\documentstyle[12pt,epsfig]{article}
        \newdimen\eqskip
        \newdimen\txtskip
        \baselineskip=\txtskip
\begin{document}
       
  \newcommand{\ccaption}[2]{
    \begin{center}
    \parbox{0.85\textwidth}{
      \caption[#1]{\small{{#2}}}
      }
    \end{center}
    }
\newcommand{\BS}{\bigskip}
\def    \be             {\begin{equation}}
\def    \ee             {\end{equation}}
\def    \ba             {\begin{eqnarray}}
\def    \ea             {\end{eqnarray}}
\def    \nn             {\nonumber}
\def    \=              {\;=\;}
\def    \frac           #1#2{{#1 \over #2}}
\def    \ret            {\\[\eqskip]}
\def    \ie             {{\em i.e.\/} }
\def    \eg             {{\em e.g.\/} }
\def \lsim{\mathrel{\vcenter
     {\hbox{$<$}\nointerlineskip\hbox{$\sim$}}}}
\def \gsim{\mathrel{\vcenter
     {\hbox{$>$}\nointerlineskip\hbox{$\sim$}}}}
\def    \bentarrow      {\:\raisebox{1.1ex}{\rlap{$\vert$}}\!\rightarrow}
\def    \rd             {{\mathrm d}}    
\def    \Im             {{\mathrm{Im}}}  
\def    \bra#1          {\mbox{$\langle #1 |$}}
\def    \ket#1          {\mbox{$| #1 \rangle$}}

\def    \kev            {\mbox{$\mathrm{keV}$}}
\def    \mev            {\mbox{$\mathrm{MeV}$}}
\def    \gev            {\mbox{$\mathrm{GeV}$}}


\def    \mq             {\mbox{$m_Q$}}  
\def    \mqq            {\mbox{$m_{Q\bar Q}$}}
\def    \mqqsq          {\mbox{$m^2_{Q\bar Q}$}}
\def    \pt             {\mbox{$p_T$}}
\def    \ptsq           {\mbox{$p^2_T$}}

\def    \o              {\ifmmode {\cal{O}} \else ${\cal{O}}$ \fi}
\def    \q              {\ifmmode {\cal{Q}} \else ${\cal{Q}}$ \fi}
\def    \oo              {\ifmmode \overline{\cal{O}} \else 
                            $\overline{\cal{O}}$ \fi}
\def    \oneSzero       {\ifmmode {^1S_0} \else $^1S_0$ \fi}
\def    \threeSone      {\ifmmode {^3S_1} \else $^3S_1$ \fi}
\def    \onePone        {\ifmmode {^1P_1} \else $^1P_1$ \fi}
\def    \threePJ        {\ifmmode {^3P_J} \else $^3P_J$ \fi}
\def    \threePzero     {\ifmmode {^3P_0} \else $^3P_0$ \fi}
\def    \threePone      {\ifmmode {^3P_1} \else $^3P_1$ \fi}
\def    \threePtwo      {\ifmmode {^3P_2} \else $^3P_2$ \fi}

\def    \heightp         {\mbox{$H_8^{\prime}$}}
\def    \vevpsi         {\mbox{$\langle {\cal O}_8^{\psi}(^3S_1) \rangle$}}
\def    \vevpsp         {\mbox{$\langle {\cal O}_8^{\psp}(^3S_1) \rangle$}}
\newcommand     \MSB            {\ifmmode {\overline{\rm MS}} \else 
                                 $\overline{\rm MS}$  \fi}
\def    \muf            {\mbox{$\mu_{\rm F}$}}
\def    \mug            {\mbox{$\mu_\gamma$}}
\def    \mufsq          {\mbox{$\mu^2_{\rm F}$}}
\def    \mur            {{\mbox{$\mu_{\rm R}$}}}
\def    \mursq          {\mbox{$\mu^2_{\rm R}$}}
\def    \mul            {{\mu_\Lambda}}
\def    \mulsq          {\mbox{$\mu^2_\Lambda$}}

\def    \as             {\mbox{$\alpha_s$}}
\def    \asb            {\mbox{$\alpha_s^{(b)}$}}
\def    \assq           {\mbox{$\alpha_s^2$}}
\def    \ech            {\mbox{$e^2_Q$ }}
\def    \bzero          {\ifmmode b_0 \else $b_0$ \fi}

\def    \eps            {\ifmmode \epsilon \else $\epsilon$ \fi}
\def    \epsbar         {\ifmmode \bar\epsilon \else $\bar\epsilon$ \fi}
\def    \epsir          {\ifmmode \epsilon_{\rm IR} \else $\epsilon_{\rm IR}$ \fi}
\def    \epsuv          {\ifmmode \epsilon_{\rm UV} \else $\epsilon_{\rm UV}$ \fi}


\def    \Atot           {{\rm A_{tot}}}
\def    \aeight         {A^{[8]}}
\def    \aeightb        {A^{[8]}_{\rm Born}}
\def    \aeights        {A^{[8]}_{\rm soft}}
\def    \ovaeight         {\overline{A^{[8]}}}
\def    \ovaeightb        {\overline{A^{[8]}}_{\rm Born}}
\def    \ovaeights        {\overline{A^{[8]}}_{\rm soft}}
\def    \aone           {A^{[1]}}
\def    \asins          {A^{[1]}_{\rm soft}}
\def    \aoneb          {A^{[1]}_{\rm Born}[^3P_J]}
\def    \aones          {A^{[1]}_{\rm soft}[^3P_J]}
\def    \aoneo          {A^{[8]}_{\rm soft}[^3P_J]}
\def    \ovaone           {\overline{A^{[1]}}}
\def    \ovaoneb          {\overline{A^{[1]}}_{\rm Born}[^3P_J]}
\def    \ovaones          {\overline{A^{[1]}}_{\rm soft}[^3P_J]}
\def    \ovaoneo          {\overline{A^{[8]}}_{\rm soft}[^3P_J]}

\def    \C              {\frac{\Phi_{(2)}}{2 M} \, \frac{N}{K}}
\def    \Cggg           {\frac{1}{2M} \frac{\Phi_{(2)}}{3!}\frac{N}{K}}
\def    \Cgggamma       {\frac{1}{2M} \frac{\Phi_{(2)}}{2!}\frac{N}{K}}
\def    \m              {\mbox{${\cal{M}}$}}
\def    \mbar           {\mbox{${\overline{\cal{M}}}$}}
\def    \mborn          {\mbox{${\cal{M}}_{\rm Born}$}}
\def    \gborn          {\mbox{$\Gamma_{\rm Born}$}}
\def    \gbh            {\mbox{$\Gamma_{\rm Born}^{H}$}}
\def    \sborn          {\mbox{$\sigma_{\rm Born}$}}
\def    \sborno         {\mbox{$\sigma^0_{\rm Born}$}}
\def    \sborno         {\mbox{$\sigma_0$}}
\def    \pgg#1          {P_{gg}(#1)}
\def    \cpgq#1         {{\cal{P}}_{gq}(#1)}
\def    \cpgg#1         {{\cal{P}}_{gg}(#1)}
\def    \cpqq#1         {{\cal{P}}_{qq}(#1)}
\def    \cpqgamma#1     {{\cal{P}}_{q\gamma}(#1)}
\def    \dk             {\mbox{${\cal D}_k$}}
\def    \fk             {\mbox{$f_k$}}
\def    \sp#1#2         {#1#2}       
\def    \eik#1          { \frac{#1 \epsilon_c}{#1 k} }
\def    \hone   {\mbox{$H_1$}}
\def    \height {\mbox{$H_8$}}

\def\jpsi{\mbox{$J\!/\!\psi$}}
\def\chic{\mbox{$\chi_c$}}
\def\chij{\mbox{$\chi_J$}}
\def\chicj{\mbox{$\chi_{cJ}$}}
\def\psp {\mbox{$\psi'$}}
\def\ups {\mbox{$\Upsilon$}}
\def\mups {\mbox{$M_\Upsilon$}}

\def \oacube {\mbox{$ O(\alpha_s^3)$}}
\def \oatwo {\mbox{$ O(\alpha_s^2)$}}
\def \oatwoa {\mbox{$ O(\alpha_s^2\aem)$}}
\def \oas   {\mbox{$ O(\alpha_s)$}}

\def \chiz {\mbox{$\chi_{0}$}}
\def \chio {\mbox{$\chi_{1}$}}
\def \chit {\mbox{$\chi_{2}$}}

\def \chiqz {\mbox{$\chi_{Q0}$}}
\def \chiqo {\mbox{$\chi_{Q1}$}}
\def \chiqt {\mbox{$\chi_{Q2}$}}

\def \chicz {\mbox{$\chi_{c0}$}}
\def \chico {\mbox{$\chi_{c1}$}}
\def \chict {\mbox{$\chi_{c2}$}}

\def \chibz {\mbox{$\chi_{b0}$}}
\def \chibo {\mbox{$\chi_{b1}$}}
\def \chibt {\mbox{$\chi_{b2}$}}

\def \QQ {Q \overline Q}
\def \qq {\mbox{$q \overline q$}}
\def \cc {\mbox{$c \overline c$}}

\def\lqcd{\mbox{$\Lambda_{QCD}$}}

\def \chizgg {\mbox{$\Gamma(\chiz \to \gamma\gamma)$}}
\def \chitgg {\mbox{$\Gamma(\chit \to \gamma\gamma)$}}
\def \chijgg {\mbox{$\Gamma(\chi_J \to \gamma\gamma)$}}
\def \chizlh {\mbox{$\Gamma(\chiz \to LH)$}}
\def \chiolh {\mbox{$\Gamma(\chio \to LH)$}}
\def \chitolh {\mbox{$\Gamma(\chij \to LH)$}}
\def \chijlh {\mbox{$\Gamma(\chi_J \to LH)$}}
\def \chijqqg{\mbox{$\Gamma(\chi_J \to q \overline q g)$}}
\def \lh{{\mathrm LH}}

\def \ebind {\mbox{$E_{\mathrm bind}$}}
\def \rprime {\mbox{${\cal R}^{\prime}_{1P}(0)$}}
\def \rprimes {\mbox{$\vert {\cal R}^{\prime}_{1P}(0) \vert^2$}}
\def \rr {\mbox{${\cal R}_S(0)$}}
\def \rros {\mbox{$\vert{\cal R}_{1S}(0)\vert^2$}}
\def \rrts {\mbox{$\vert{\cal R}_{2S}(0)\vert^2$}}
\def \rrns {\mbox{$\vert{\cal R}_{nS}(0)\vert^2$}}
\def \ei {\mbox{$ \epsilon_{ \!\!\mbox{ \tiny{IR} } } $}}
\def \eu {\mbox{$ \epsilon_{ \!\!\mbox{ \tiny{UV} } } $}}
\def \s0 {\mbox{$\sigma_{0} $}}
\def \se {\mbox{$\sigma_{0}(\epsilon) $}}
\def \ep {\mbox{$\epsilon $}}

\def \cf {\mbox{$ C_F $}}
\def \ca {\mbox{$ C_A $}}
\def \caf {\mbox{$ C_F-\frac{1}{2}C_A $}}
\def \tf {\mbox{$ T_F $}}
\def \nf {\mbox{$n_f$}}
\def \da {\mbox{$ D_A $}}
\def \Bf {\mbox{$ B_F $}}
\def \df {\mbox{$ D_F $}}

\def \fe{\mbox{$f(\ep)$}}
\def \de{\mbox{$F(\ep)$}}
\def \feps#1 {f_{\epsilon}(#1)}

\def\der{\mbox{$\stackrel{\leftrightarrow}{\bf D}$}}
\def\nder{\mbox{$\stackrel{\leftrightarrow}{D}$}}
\def\tr{{\mathrm Tr}}

\def\opchizh {\mbox{$\langle 0\vert {\cal O}_8^{\psi}(^3P_0)\vert 0 \rangle$}}
\def\opchioh {\mbox{$\langle 0\vert {\cal O}_8^{\psi}(^3P_1)\vert 0 \rangle$}}
\def\opchith {\mbox{$\langle 0\vert {\cal O}_8^{\psi}(^3P_2)\vert 0 \rangle$}}

\def\opchijh {\mbox{$\langle 0\vert {\cal O}_8^{\psi}(^3P_J)\vert 0 \rangle$}}
\def\opetah  {\mbox{$\langle 0\vert {\cal O}_8^{\psi}(^1S_0)\vert 0 \rangle$}}
\def\oppsih  {\mbox{$\langle 0\vert {\cal O}_8^{\psi}(^3S_1)\vert 0 \rangle$}}
\def\oppsis  {\mbox{$\langle 0\vert {\cal O}_1^{\psi}(^3S_1)\vert 0 \rangle$}}
\def\pppsisq  {\mbox{$\langle 0\vert {\cal P}_1^{\psi_Q}(^3S_1)\vert 0 \rangle$}}

\def\opchijhq {\mbox{$\langle 0\vert {\cal O}_8^{\psi_Q}(^3P_J)\vert 0 \rangle$}}
\def\opetahq  {\mbox{$\langle 0\vert {\cal O}_8^{\psi_Q}(^1S_0)\vert 0 \rangle$}}
\def\oppsihq  {\mbox{$\langle 0\vert {\cal O}_8^{\psi_Q}(^3S_1)\vert 0 \rangle$}}
\def\oppsisq  {\mbox{$\langle 0\vert {\cal O}_1^{\psi_Q}(^3S_1)\vert 0 \rangle$}}
\def\ppchijh  {\mbox{$\langle 0\vert {\cal O}_8^{\tiny\chij}(^3S_1)\vert 0 \rangle$}}
\def\opchizhq {\mbox{$\langle 0\vert {\cal O}_8^{\psi_Q}(^3P_0)\vert 0 \rangle$}}
\def\opchiohq {\mbox{$\langle 0\vert {\cal O}_8^{\psi_Q}(^3P_1)\vert 0 \rangle$}}
\def\opchithq {\mbox{$\langle 0\vert {\cal O}_8^{\psi_Q}(^3P_2)\vert 0 \rangle$}}

\def\ppchizh  {\mbox{$\langle 0\vert {\cal O}_8^{\tiny\chiz}(^3S_1)\vert 0 \rangle$}}
\def\ppchioh  {\mbox{$\langle 0\vert {\cal O}_8^{\tiny\chio}(^3S_1)\vert 0 \rangle$}}
\def\ppchith  {\mbox{$\langle 0\vert {\cal O}_8^{\tiny\chit}(^3S_1)\vert 0 \rangle$}}

\def\ppchiczh  {\mbox{$\langle 0\vert {\cal O}_8^{\tiny\chicz}(^3S_1)\vert 0 \rangle$}}
\def\ppchicoh  {\mbox{$\langle 0\vert {\cal O}_8^{\tiny\chico}(^3S_1)\vert 0 \rangle$}}
\def\ppchicth  {\mbox{$\langle 0\vert {\cal O}_8^{\tiny\chict}(^3S_1)\vert 0 \rangle$}}

\def\oppsph  {\mbox{$\langle 0\vert {\cal O}_8^{\psi^{\prime}}(^3S_1)\vert 0 \rangle$}}
\def\oppsps  {\mbox{$\langle 0\vert {\cal O}_1^{\psi^{\prime}}(^3S_1)\vert 0 \rangle$}}

\def\spectrc {\mbox{$^{2S+1}L^{[c]}_J$}}
\def\spectr {\mbox{$^{2S+1}L_J$}}
\def\spectrs {\mbox{$^{2S+1}L_J^{[1]}$}}
\def\spectrh {\mbox{$^{2S+1}L_J^{[8]}$}}
\def\szh    {\mbox{$\sigma_0^{H}$}}
\def\sbh    {\mbox{$\sigma_{\rm Born}^{H}$}}
\def\szpsi {\mbox{$\sigma_0^{\psi}$}}
\def\szpsiq {\mbox{$\sigma_0^{\psi_Q}$}}
\def\spectro {{^{2S+1}L^{[8]}_J}}

\def\etah {\ifmmode {^1S_0^{[8]}} \else  {$^1S_0^{[8]}$} \fi}
\def\etas {\ifmmode {^1S_0^{[1]}}  \else  {$^1S_0^{[1]}$} \fi}
\def\psih {\ifmmode {^3S_1^{[8]}}  \else  {$^3S_1^{[8]}$} \fi}
\def\psis {\ifmmode {^3S_1^{[1]}}  \else  {$^3S_1^{[1]}$} \fi} 
\def\chizh {\ifmmode {^3P_0^{[8]}}  \else  {$^3P_0^{[8]}$} \fi}
\def\chioh {\ifmmode {^3P_1^{[8]}}  \else  {$^3P_1^{[8]}$} \fi}
\def\chith {\ifmmode {^3P_2^{[8]}}   \else  {$^3P_2^{[8]}$} \fi}
\def\chijh {\ifmmode {^3P_J^{[8]}}  \else  {$^3P_J^{[8]}$} \fi} 
\def\chizs {\mbox{$^3P_0^{[1]}$}}
\def\chios {\mbox{$^3P_1^{[1]}$}}
\def\chits {\mbox{$^3P_2^{[1]}$}}
\def\chijs {\mbox{$^3P_J^{[1]}$}}
\def\chijo {\mbox{$^3P_J^{[8]}$}}
\def\opchizs {\mbox{$\langle 0\vert {\cal O}_1^{\tiny\chiz}(^3P_0)\vert 0 \rangle$}}
\def\opchios {\mbox{$\langle 0\vert {\cal O}_1^{\tiny\chio}(^3P_1)\vert 0 \rangle$}}
\def\opchits {\mbox{$\langle 0\vert {\cal O}_1^{\tiny\chit}(^3P_2)\vert 0 \rangle$}}
\def\opchijs {\mbox{$\langle 0\vert {\cal O}_1^{\tiny\chij}(^3P_J)\vert 0 \rangle$}}
\def\opchizh {\mbox{$\langle 0\vert {\cal O}_8^{\psi}(^3P_0)\vert 0 \rangle$}}
\def\opchioh {\mbox{$\langle 0\vert {\cal O}_8^{\psi}(^3P_1)\vert 0 \rangle$}}
\def\opchith {\mbox{$\langle 0\vert {\cal O}_8^{\psi}(^3P_2)\vert 0 \rangle$}}
\def\opchijh {\mbox{$\langle 0\vert {\cal O}_8^{\psi}(^3P_J)\vert 0 \rangle$}}
\def\opetah  {\mbox{$\langle 0\vert {\cal O}_8^{\psi}(^1S_0)\vert 0 \rangle$}}
\def\oppsih  {\mbox{$\langle 0\vert {\cal O}_8^{\psi}(^3S_1)\vert 0 \rangle$}}
\def\oppsis  {\mbox{$\langle 0\vert {\cal O}_1^{\psi}(^3S_1)\vert 0 \rangle$}}
\def\ophtpzs {\mbox{$\langle 0\vert {\cal O}_1^{H}(^3P_0)\vert 0 \rangle$}}
\def\ophtpos {\mbox{$\langle 0\vert {\cal O}_1^{H}(^3P_1)\vert 0 \rangle$}}
\def\ophtpts {\mbox{$\langle 0\vert {\cal O}_1^{H}(^3P_2)\vert 0 \rangle$}}
\def\ophtpjs {\mbox{$\langle 0\vert {\cal O}_1^{H}(^3P_J)\vert 0 \rangle$}}
\def\ophoszs {\mbox{$\langle 0\vert {\cal O}_1^{H}(^1S_0)\vert 0 \rangle$}}
\def\ophtsos {\mbox{$\langle 0\vert {\cal O}_1^{H}(^3S_1)\vert 0 \rangle$}}
\def\ophtpzo {\mbox{$\langle 0\vert {\cal O}_8^{H}(^3P_0)\vert 0 \rangle$}}
\def\ophtpoo {\mbox{$\langle 0\vert {\cal O}_8^{H}(^3P_1)\vert 0 \rangle$}}
\def\ophtpto {\mbox{$\langle 0\vert {\cal O}_8^{H}(^3P_2)\vert 0 \rangle$}}
\def\ophtpjo {\mbox{$\langle 0\vert {\cal O}_8^{H}(^3P_J)\vert 0 \rangle$}}
\def\ophoszo {\mbox{$\langle 0\vert {\cal O}_8^{H}(^1S_0)\vert 0 \rangle$}}
\def\ophtsoo {\mbox{$\langle 0\vert {\cal O}_8^{H}(^3S_1)\vert 0 \rangle$}}

\def\opchih {\mbox{$\langle 0\vert{\cal O}_8^{\tiny\chij}(^3S_1)\vert 0\rangle$}}

\def\b0{\mbox{$b_0$}}
\def\dsh  {\mbox{$\sigma^{H} $}}  
\def\dspq  {\mbox{$d\sigma^{\psi_Q} $}}
\def\dsp  {\mbox{$d\sigma^{\psi_Q} $}}
\def\qf     {\mbox{$\mu^2_{\rm F}$}}
\def\asopi{\mbox{$\frac{\as}{\pi}$}}
\def\szchiz{\mbox{$\sigma_0^{\tiny\chiz}$}}
\def\szchio{\mbox{$\sigma_0^{\tiny\chio}$}}
\def\szchit{\mbox{$\sigma_0^{\tiny\chit}$}}
\def\szchij{\mbox{$\sigma_0^{\tiny\chij}$}}
\def\dschiz  {\mbox{$d\sigma^{\chiz} $}}
\def\dschio  {\mbox{$d\sigma^{\tiny\chio} $}}
\def\dschit  {\mbox{$d\sigma^{\tiny\chit} $}}
\def\dschij  {\mbox{$d\sigma^{\tiny\chij} $}}

\def\vd{\mbox{${\bf D}$}}
\def\ve{\mbox{${\bf E}$}}
\def\vb{\mbox{${\bf B}$}}
\def\vs{\mbox{${\bf \sigma}$}}

\def\rj {{\mathrm J}}

\def \chizgg {\mbox{$\Gamma(\chiz \to \gamma\gamma)$}}
\def \chitgg {\mbox{$\Gamma(\chit \to \gamma\gamma)$}}
\def \chijgg {\mbox{$\Gamma(\chi_J \to \gamma\gamma)$}}
\def \chizlh {\mbox{$\Gamma(\chiz \to \lh)$}}
\def \chiolh {\mbox{$\Gamma(\chio \to \lh)$}}
\def \chitlh {\mbox{$\Gamma(\chit \to \lh)$}}
\def \chijlh {\mbox{$\Gamma(\chi_J \to \lh)$}}

\def\aem{\mbox{$\alpha_{{\mathrm\tiny em}}$}}
\def\aemb{\mbox{$\alpha^{(b)}_{{\mathrm\tiny em}}$}}
\def\coll{\mbox{$ \vert\vert$}}
\def\Qb{\mbox{$\overline Q$}}
\def\ko{\mbox{$k_1$}}
\def\kt{\mbox{$k_2$}}
\def\mqs{\mbox{$m_Q^2$}}
\def\ks{\mbox{$k^2$}}
\def\ne{\mbox{$N_\epsilon$}}
\def\mzs{\mbox{$m_0^2$}}
\def\mos{\mbox{$m_1^2$}}
\def\mds{\mbox{$m_2^2$}}
\def\mts{\mbox{$m_3^2$}}
\def\mfs{\mbox{$m_4^2$}}
\def\tu{\mbox{$t_1$}}
\def\uu{\mbox{$u_1$}}
\def\li{\mbox{$\mathrm{Li}_2$}}
\def\meas{\mbox{$\frac{\rd^D k}{(2\pi)^D}$}}
\def\fc{\mbox{${\mathrm F} \chi$}}
\def\nfc{\mbox{${\mathrm NF} \chi$}}

\def\mufrag{\mbox{$\mu_{\mathrm frag}$}}
\def\mufrags{\mbox{$\mu^2_{\mathrm frag}$}}

\def\mct{\mbox{$m_c^3$}}
\def\mcf{\mbox{$m_c^5$}}

\def\msb{\mbox{${\overline{MS}}$}}
\def\shad{\mbox{$S_{\mathrm had}$}}
\def\nc{\mbox{$N_c$}}
\def\psiq{\mbox{$\psi_Q$}}
\def\pspq{\mbox{$\psi^\prime_Q$}}
\def\chijq{\mbox{$\chi_{QJ}$}}
\def\nj{\mbox{$N_J^\epsilon$}}

\def\bfp{{\bf p}}
\def\bfpp{{\bf p}'}
\def\bfk{{\bf k}}

\def\slash#1{{#1\!\!\!/}}

\begin{titlepage}
\nopagebreak
{\flushright{
        \begin{minipage}{5cm}
        CERN-TH/97-202\\
        hep-ph/9708349\\
        \end{minipage}        }

}
\vfill
\begin{center}
{\LARGE { \bf \sc Quarkonium Photoproduction \\[0.5cm]
at Next-to-leading Order}}
\vfill                                                       
\vskip .5cm
{\bf Fabio MALTONI\footnote{Permanent address:
     Dipartimento di Fisica dell'Universit\`{a} and Sez. INFN, Pisa, Italy} ,
Michelangelo L. MANGANO\footnote{On leave of absence from                    
    INFN, Pisa, Italy}}\\
{CERN, TH Division, Geneva, Switzerland} \\
\verb+fabio.maltoni@cern.ch, mlm@vxcern.cern.ch+\\
\vskip .5cm
and {\bf Andrea PETRELLI\footnote{Address after Oct. 1st: Theory Division, 
Argonne National Laboratory, Argonne, IL, USA.} } \\
{INFN, Sezione di Pisa, Italy} \\                      
\verb+petrelli@ibmth.difi.unipi.it+\\
\end{center}
\nopagebreak
\vfill
\begin{abstract} 
We present the calculation of $\oatwoa$ corrections to heavy-quarkonium total
photoproduction cross-sections. Results are given for 
the colour-octet component of $S$ and $P$
waves. The calculation is performed using
covariant projectors in dimensional regularization. A phenomenological study of
the results, including a discussion of the high-energy behaviour of the cross
sections, is presented. For $\gamma p$ energies up to few hundred GeV
the NLO corrections significantly reduce the scale          
dependence of the production rates relative to the Born-level results.
Large small-$x$ corrections arise at higher energies, making the predictions
strongly dependent on the shape of the gluon density and on the choice of
factorization scale.
\end{abstract}                                                
\vskip 1cm
CERN-TH/97-202\hfill \\
\today \hfill
\vfill 
\end{titlepage}
\section{Introduction}
The calculation of production and decay processes for heavy quarkonium states
has recently been put on a solid formal basis by the work of Bodwin, Braaten
and Lepage~\cite{bbl} (BBL).                  
According to their results, production and decay rates can be
calculated within perturbative QCD as the sum of products of short-distance
coefficients times long-distance matrix elements.
The short-distance coefficients are the square of transition matrix elements 
for production or decay of heavy-quark pairs in definite states of colour, spin
and angular momentum.
The long-distance ones are obtained from the matrix elements
of quark-antiquark operators with the same quantum numbers as those of the
short-distance state; these are 
evaluated between the vacuum and an arbitrary state
containing the physical quarkonium meson we are interested in, 
squared and summed over all possible final states accompanying the quarkonium.
These long-distance                               
matrix elements can in principle be calculated on the lattice, and hierarchies
among them can be obtained by applying the velocity scaling rules of 
NRQCD~\cite{CL86,LMNMH92}.
Several applications of   
this formalism have been obtained, and are nicely reviewed in
ref.~\cite{Braaten96a,Beneke97a}.

One of the most important consequences of this factorization
property of quarkonium
production is the prediction that the value of the non-perturbative parameters
does not depend on the details of the hard process, so that parameters
extracted from a given experiment can be used in different ones. For
simplicity, we will refer to this concept as ``universality''. 
Several studies of experimental data coming 
from different kind of reactions have been performed to assess the validity
of universality. For example  calculations of inclusive quarkonia production
in $e^+e^-$ annihilation~\cite{ee}, fixed target experiments~\cite{coft}, 
$\gamma p$  collisions~\cite{Cacciari96,gammap} and $B$ decays 
\cite{bdecays} have  been carried out within this framework. 
The overall agreement of theory and data is satisfactory, but there are clear
indications that large uncertainties are present. The most obvious one is the
discrepancy~\cite{Cacciari96}
between  HERA data~\cite{HERApsi} and the large amount  of
inelastic \jpsi\ photoproduction predicted by applying the colour-octet matrix
elements extracted~\cite{Cacciari95,Cho96,Beneke97} from the Tevatron
large-\pt\  data~\cite{cdf,d0}.
                               
In view of this discrepancy,
it becomes important to assess to which extent is universality
applicable. Several potential sources of universality violation are indeed
present, both at the perturbative and non-perturbative level.
On one hand there are potentially large corrections to the factorization
theorem itself. In the case of charmonium production, for example, the mass of
the heavy quark is small enough that non-universal power-suppressed corrections
can be large.  Furthermore, some higher-order corrections in the velocity
expansion are strongly enhanced at the edge of phase-space~\cite{Beneke97b}.
For example, the alternative choices of using as a mass parameter for the
matrix elements and for the phase-space boundary
the mass of a given quarkonium state or twice the heavy-quark mass $2m$,
give rise to a large uncertainty in the production rate near threshold.
These effects, which are present both in the total cross-section and in the
production at large-\pt\ via gluon fragmentation, violate universality. This is
because the threshold behaviour depends on the nature of the hard process under
consideration.

Another source of bias in the use of universality comes from purely
perturbative corrections. Most of the current predictions for quarkonium
production are based on the use of leading-order (LO) matrix elements. Possible
perturbative $K$-factors are therefore absorbed into the non-perturbative matrix
elements extracted from the comparison of data with theory. Since the size of
the perturbative corrections varies from one process to the other, an
artificial violation of universality is introduced. Examples of the size of
these corrections are given by the large impact of $k_T$-kick effects and
initial-state multiple-gluon emission in open-charm~\cite{Frixione97} and
charmonium production~\cite{e789,CanoColoma97}.

In ref.~\cite{Petrelli97} we focused on the evaluation of the \oacube\
corrections to quarkonium total hadro-production cross-sections.  As pointed
out in ref.~\cite{Mangano96},  the impact of NLO corrections can be significant
and a general study of their effects is necessary. In this paper we concentrate
on the calculation of the $\oatwoa$ corrections to total
photoproduction cross-sections. To carry out our calculations, we need a
framework for calculating NLO inclusive production cross-sections. As well
known, the most convenient method for  regulating both UV and
IR divergences in perturbative calculations 
beyond leading order in  $\as$ is dimensional regularization. On  the other 
hand most calculations of production cross-sections and decay rates for  heavy
quarkonia 
have been performed using the covariant projection method \cite{Kuehn79}, 
which involves the  projection of the $\QQ$ pair onto states with definite 
total angular momentum $J$, and which is specific to four dimensions.

In ref.~\cite{Petrelli97} we presented a generalization of the
method of covariant projection  to $D = 4 - 2\epsilon$ dimensions. In that
paper our formalism was shown to provide equal results to calculations
performed within the ``threshold expansion'' technique, introduced 
in~\cite{Braaten96,Braaten97}.
In this work we apply the covariant-projection technique 
to the calculation of the $\oatwoa$ total
cross-sections for the photoproduction of several $\QQ$ states of  
phenomenological relevance: $\oneSzero^{[8]}$, $\threeSone^{[8]}$ and
$\threePJ^{[8]}$, where the right upper index labels the colour configuration
of the $\QQ$ pair. 
                                         
The paper is structured as follows. In Section~\ref{sec:projectors} we 
briefly review our formalism. A more complete discussion can be found in
ref.~\cite{Petrelli97}.        
Section~\ref{sec:soft} gives a brief general description of the 
NLO calculation. In particular, we describe the 
the behaviour of the soft 
limit of the NLO real corrections and the technique used to identify the
residues of the IR and collinear singularities and to allow their cancellation
without the need for a complete $D$-dimensional calculation of the
real-emission matrix elements.
Section~\ref{sec:production} 
presents the various components (real and virtual corrections) of the
NLO calculation of the production processes. 
Section~\ref{sec:pheno} presents a numerical study of the results, with a
discussion of the individual components of the cross-sections, of the $K$
factors, and of the scale dependence.
A discussion of the high-energy behaviour of the production rates at
NLO and of the uncertainties of the predictions in this regime is included.
The last section contains our conclusions, as well as a discussion of the
relevance of our calculation for the study of photoproduction at $\pt>0$ and
$z<1$.
     
Appendix~\ref{appA} collects symbols and notations. A summary of all 
results is provided in Appendix~\ref{appNLO}, where the parton-level
photoproduction cross-sections are presented in  their final
form, after the cancellation of all singularities.

\section{Introduction to the formalism}
\label{sec:projectors}
The starting point of the calculation of the production cross-sections is given
by the standard formula~\cite{bbl}:
\be                                              
d\sigma(H + X) = \sum_n d\hat\sigma(\QQ [n] + X)\langle{\cal O}^H(n)\rangle
\label{eq:xsect}                                
\ee
The quantity $\langle{\cal O}^H(n)\rangle$ is proportional to the 
inclusive transition probability of the perturbative quark-pair
state $\QQ[n]$, with quantum numbers labelled by $n$, into the
quarkonium  state $H$. $d\hat\sigma(\QQ [n] + X)$ is the 
short-distance cross-section for the production of the
perturbative state $\QQ[n]$.
It can be calculated in perturbative QCD                                       
either using threshold-mathing techniques~\cite{bbl,Braaten96}, or using
projection techniques~\cite{Kuehn79}.                                           
These projection techniques have been recently extended for use in
$D$-dimensions~\cite{Petrelli97}, to deal with
the presence of infrared (IR) and ultraviolet (UV) divergences. 
They will be briefly reviewed here.
     
The spin projectors, with non-relativistic normalization for the spinors, for
outgoing heavy quarks momenta $Q = P/2+q$ and ${\overline Q }= P/2-q$, are 
given by~\cite{Kuehn79}:
\ba                                                                
&&{\mit \Pi}_{0}
     = {1\over{\sqrt{8m^3}}} \left({{\slash{P}}\over 2} - \slash{q} - m\right)
       \gamma_5 \left({{\slash{P}}\over 2} + \slash{q} + m\right)\, ,
\label{proj_00}
\\
&&{\mit \Pi}_{1}^{\alpha} 
     = {1\over{\sqrt{8m^3}}} \left({{\slash{P}}\over 2} - \slash{q} - m\right)
       \gamma^\alpha \left({{\slash{P}}\over 2} + \slash{q} + m\right)\, ,
\label{proj_1Sz}
\ea
for spin zero and spin one states respectively. In these relations, $P$ is the
momentum of the quarkonium state, $2q$ is the relative momentum between the
$\QQ$ pair, and $m\equiv M/2$ is the mass of the heavy quark $Q$. 
The justification for use of these projectors in $D$ dimensions, and a
discussion of how to deal with the presence of the $\gamma_5$ matrix, can be
found in ref.~\cite{Petrelli97}. Here we limit ourselves to pointing out that
the $D$-dimensional character of space-time is implicit in
eqs.~(\ref{proj_00},\ref{proj_1Sz}), and appears explicitly when performing the
sums over polarizations, as shown later.

The colour singlet or octet state content of a given state will  be
projected out by contracting the amplitudes with the following
operators :
\ba
&&{\cal C}_1 = {{\delta_{ij}}\over{\sqrt{\nc}}}\qquad{\rm for~the~singlet}
\label{proj_sing}
\\
&&{\cal C}_8 = \sqrt{2} T_{ij}^c\qquad{\rm for~the~octet}
\label{proj_oct}
\ea

The projection on a state with orbital angular momentum $L$ is obtained by 
differentiating $L$ times the spin- and colour-projected amplitude with respect
to  the momentum $q$ of the heavy quark in the $\QQ$ rest frame, and then 
setting $q$ to zero. We shall only
deal with either $L=0$ or $L=1$ states, for which the amplitudes take
the form:
\ba      
&&{\cal A}_{S=0,L=0} = \tr\left.\left[{\cal C}\,{\mit  \Pi}_0\, {\cal A}\right]
\right|_{q=0}\qquad\qquad\qquad\,{\rm Spin}\;{\rm  singlet }~S~{\rm states}\\
&&{\cal A}_{S=1,L=0} = \tr\left.\left[{\cal C}\,{\mit  \Pi}_1^\alpha\, 
{\cal A}\right]\right|_{q=0}\epsilon_\alpha\qquad\qquad\;\;\,\,\,{\rm Spin} 
\;{\rm triplet }~S~{\rm states}\\                                        
&&{\cal A}_{S=0,L=1} = {{\rd}\over{\rd q_\beta}} \tr\left.\left[{\cal C}\, 
{\mit \Pi}_0 
{\cal A}\right]\right|_{q=0}\epsilon_\beta\quad\qquad\;\;\,{\rm Spin}\;
{\rm singlet }~P~{\rm states}\\
&&{\cal A}_{S=1,L=1} = {{\rd}\over{\rd q_\beta}} \tr\left.\left[{\cal C}\, 
{\mit \Pi}_1^\alpha 
{\cal A}\right]\right|_{q=0}{\cal E}_{\alpha\beta}\quad\qquad{\rm Spin}\; 
{\rm triplet }~P~{\rm states}
\ea    
${\cal A}$ being the standard QCD amplitude for production (or decay) of the 
heavy quark and antiquark $Q$ and ${\overline Q}$, amputated of the heavy quark
spinors.

The amplitudes ${\cal A}_{S,L}$ will then have to be squared, summed over the
final degrees of freedom and averaged over the initial ones.
                                                            
The selection of the appropriate total angular momentum quantum number is done
by performing the proper polarization sum. We define:
\be                                                  
\Pi_{\alpha\beta} \equiv -g_{\alpha\beta} + {{P_\alpha P_\beta}\over{M^2}}
\; .
\ee
The sum over polarizations for a $^3S_1$ state, which is still a vector even in
$D=4-2\ep$ dimensions, is then given by:
\be                       
\sum_{J_z} \ep_\alpha\ep_{\alpha'}^* = \Pi_{\alpha\alpha'}
\ee
In the case of $^3P_J$ states, the three multiplets corresponding to $J=0,1$
and 2 correspond to a scalar, an antisymmetric tensor and a symmetric
traceless tensor, respectively. We shall denote their polarization tensors by 
${\cal E}_{\alpha\beta}^{(J)}$. The sum over polarizations is then given by:
\ba                                                           
{\cal E}_{\alpha\beta}^{(0)}{\cal E}_{\alpha'\beta'}^{(0)*} &=&
{1\over{D-1}} \Pi_{\alpha\beta}\Pi_{\alpha'\beta'}
\label{polsum0}
\\
\sum_{J_z} {\cal E}_{\alpha\beta}^{(1)}{\cal E}_{\alpha'\beta'}^{(1)*} &=&
{1\over{2}} [\Pi_{\alpha\alpha'}\Pi_{\beta\beta'} - 
             \Pi_{\alpha\beta'}\Pi_{\alpha'\beta}]
\label{polsum1}
\\
\sum_{J_z} {\cal E}_{\alpha\beta}^{(2)}{\cal E}_{\alpha'\beta'}^{(2)*} &=&
{1\over{2}} [\Pi_{\alpha\alpha'}\Pi_{\beta\beta'} +                       
             \Pi_{\alpha\beta'}\Pi_{\alpha'\beta}] - 
{1\over{D-1}} \Pi_{\alpha\beta}\Pi_{\alpha'\beta'}
\label{polsum2}
\ea
for the $\threePzero$, $\threePone$ and $\threePtwo$ states respectively.
Total contraction of the polarization tensors gives the number of polarization
degrees of freedom in $D$ dimensions. Therefore
\be
N_J = \sum_{J_z} \ep_\alpha\ep_{\alpha}^* = \Pi_{\alpha\alpha} = D-1 = 3 -2 \ep
\ee
for the $\threeSone$ state and
\be
N_J = \sum_{J_z} {\cal E}_{\alpha\beta}^{(J)}{\cal E}_{\alpha\beta}^{(J)*}
\ee
for the $\threePJ$  states, with
\be                 
N_0 = 1, \; \;  N_1 = {{(D-1)(D-2)}\over 2}=(3-2\ep)(1-\ep), \; \;
N_2 = {{(D+1)(D-2)}\over 2}=(5-2\ep)(1-\ep) \;.                   
\ee                                            

The application of this set of rules produces the short-distance cross
section coefficients $\hat \sigma$ 
for the $ij\rightarrow\spectr^{[1,8]}$ processes:
\be         
d\hat\sigma(ij\to\spectr^{[1,8]}) = {1\over{2s}}
   \overline{\sum}|{\cal A}_{S,L}|^2\;d \Phi\, ,
\ee                                             
$s$ being the partonic centre of mass energy squared.
To find the physical cross-sections 
for the observable quarkonium state $H$ these short distance coefficients
must be properly related to the NRQCD production  matrix
elements $\langle{\cal O}_{[1,8]}^H(\spectr)\rangle$.
The cross-sections then read
\be                     
\sigma(ij\to\spectr^{[1,8]}\to H) = \hat \sigma(ij\to\spectr^{[1,8]}) 
{{\langle \o_{[1,8]}^H(\spectr)\rangle}
\over{N_{col} N_{pol}}} \; ,
\ee                         
where $N_{col}$ and $N_{pol}$ refer to the number of colours and polarization
states of the $\QQ[\spectr]$ pair produced. They are given by 1 for singlet
states or $\nc^2-1$ for octet states, and by the $D$-dimensional  $N_J$'s
defined above. Dividing by these colour and polarization  degrees of freedom in
the cross-sections is necessary as we had summed over them in the evaluation of
the short distance coefficient $\hat\sigma$. 
As discussed at length in ref.~\cite{Petrelli97}, our conventions for the
normalization of the non-perturbative matrix elements differ slightly from the
conventional ones introduced by BBL~\cite{bbl}:
\ba                                                             
&&\langle {\o}_1\rangle = \frac{\langle {\o }_1\rangle ^{\rm BBL}}{2\nc}, \\
&&\langle {\o}_8\rangle = \langle {\o }_8\rangle ^{\rm BBL}. 
\ea                                                          

\section{Soft factorization and the calculation of higher-order corrections}
\label{sec:soft}                
In this section we discuss the r\^{o}le played by the universal IR behaviour of
gluon-emission amplitudes in the  calculation of  higher-order corrections to
total production cross-sections.
A consistent calculation of higher-order corrections entails the evaluation of
the real and virtual emission diagrams, carried out in $D$ dimensions. The UV
divergences present in the virtual diagrams are removed by the standard
renormalization. The IR divergences appearing after the
integration over the phase space of the emitted parton are cancelled by
similar divergences present in the virtual corrections.
Collinear divergences, finally, are either cancelled by similar divergences
in the virtual corrections or by factorization
into the NLO parton densities. 
The evaluation of the real emission matrix elements in $D$ dimensions 
is usually particularly complex.
In this paper we follow an approach already employed
in~\cite{Mele91}, whereby the structure of soft and collinear singularities in
$D$ dimensions is extracted by using universal factorization properties of the
amplitudes. Thanks to these factorization properties, that will be discussed in
detail in the following section, 
the residues of IR and collinear poles in $D$ dimensions
can be obtained without an explicit calculation of the full
$D$-dimensional real matrix elements. They only require, in general,
knowledge of the $D$-dimensional Born-level amplitudes, a much simpler task.
The isolation of these residues allows to
carry out the complete cancellations of the relative poles in $D$ dimensions,
leaving residual finite expressions which can then be evaluated exactly
directly in $D=4$ dimensions. 
The four-dimensional real matrix elements that we will need
can be found in the literature~\cite{Cacciari96}.
                              
\subsection{Soft factorization in $gg \gamma $ amplitudes}
\begin{figure}[t]                                 
\begin{center}
\epsfig{file=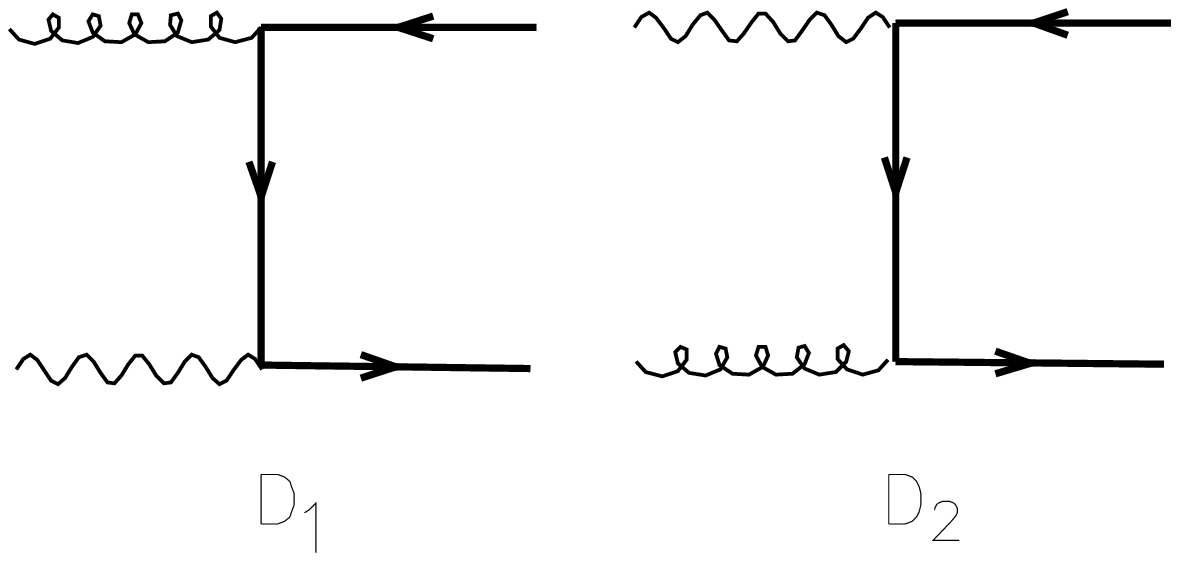,width=0.75\textwidth,clip=}
\ccaption{}{\label{fig:ggborn} 
Diagrams for the $g\gamma$ Born amplitudes.}
\end{center}
\end{figure}

At the Born level, the
relevant diagrams are shown in fig.~\ref{fig:ggborn}. The production amplitude
(before projection on a specific quarkonium state) can
be written as follows:                             
\be
  A_{\rm Born} \= T^a_{ij} \; D_{12} \; ,
\ee           
where $ D_{12} \equiv D_1+ D_2 $ and the  terms $D_1$, $D_2$  
correspond to the diagrams appearing in
fig.~\ref{fig:ggborn} with the colour coefficients removed. Using this
notation, 
the amplitude for emission of a soft gluon with momentum $k$ and
colour label $c$ can be written as follows~\cite{Bassetto83}:
\ba                                                                   
    A_{\rm soft} &=&  
   g(T^c T^a)_{ij} \left[ \eik{Q} - \eik{a} \right] D_{12}  +
   g(T^a T^c)_{ij} \left[ \eik{a} - \eik{\overline{Q}} \right]   D_{12}+
\ea                                                                      
where $Q$ and $\overline{Q}$ are the momenta of the heavy quarks, and $a$
indicates the momentum and colour label of the initial state gluon.
Born-level colour-singlet amplitudes vanish, so
we will concentrate on
colour-octet states. Using the projectors defined in the previous section,
we can write:
 \ba                                                                   
    A^{[8]}_{\rm soft} &=& g \sqrt{2} D_{12} \left\{
   \tr (T^b T^c T^a) \left[ \eik{Q} - \eik{a} \right] +
   \tr (T^b T^a T^c) \left[ \eik{a} - \eik{\overline{Q}} \right]\right\}
\nn\\
&=& g \frac{\sqrt{2}}{4}  D_{12}\left\{
    d^{bca} \left[ \eik{Q} - \eik{\overline{Q}} \right] +
    i f^{bca} \left[ \eik{Q}  + \eik{\overline{Q}} 
    - 2 \eik{a} \right]\right\}\, ,
\ea                                
where $b$ is the colour label of the colour-octet $\QQ$ state.

In the case of $S$-wave production we can set $Q=\overline{Q}=P/2$, and we get
\ba                                                         
   \left( A^{[8]}_{\rm soft}\right) _{q=0}  &=& 
    \frac{ i g f^{bca}} {\sqrt{2}}  
   \left[ \eik{P} - \eik{a} \right] \,D_{12}\, .
\ea                       
In the case of $^3P$-waves we also need the derivative of the decay
amplitude with respect to the relative momentum of the quark and antiquark.
In the soft-gluon limit, we obtain:
\be 
    \left(\frac{dA^{[8]}_{\rm soft}}{dq_{\alpha}} \right)_{q=0} \=
    g \frac{\sqrt{2}}{2} \left\{                                  
      i f^{bca}  \left[ \eik{P} - \eik{a} \right] 
      \frac{d D_{12} }{dq_{\alpha}} + d^{bca}
    \left[\frac{\epsilon_c^{\alpha}}{\sp{P}{k} }
     - \frac{k^{\alpha} }{(Pk)^2} (P \epsilon_c) \right] D_{12} \right\}
\ee                                                                      
Choosing a transverse gauge where $\epsilon_c \cdot P=0$,
from eq.~(9) and the previous two expressions it is straightforward to write
\be                                                 
{\cal A}_{^3 P_J} \= {\cal E}_{\alpha\beta} \frac{ i g f^{bca}}{\sqrt{2}}  
   \left[ \eik{P} - \eik{a} \right]           
{{\rd}\over{\rd q_\beta}} \tr\left. \left[{\cal C}\,
{\mit \Pi}_1^\alpha
{\cal A}_{\rm Born} \right]\right|_{q=0} 
 \;+\; \frac{g d^{bca}}{\sqrt{2}\sp{P}{k} } {\cal E}_{\alpha\beta}
\epsilon_c^{\alpha} \tr\left. \left[{\cal C}\,
{\mit \Pi}_1^\alpha
{\cal A}_{\rm Born} \right]\right|_{q=0}  \; .
\label{eq:dAdq}                               
\ee
The soft amplitude factorizes if the last term vanishes.
This last term can be seen to be proportional to 
the  amplitude for the production of a                                 
$^3S_1$ state, with an {\em effective} polarization $\epsilon_{\rm eff}$
given in terms of the polarizations of the 
soft gluon and of the $^3P_J$ state as follows:
\be        
   \epsilon^{J}_{{\rm eff}, \beta} \=
   \epsilon_c^{\alpha} {\cal E}_{\alpha\beta}^{J}(P)  \; .
\ee                                                     
If the average on the soft gluon $D-1$ spatial 
directions is taken, \ie
\be
 \int \frac{d\Omega^{D-1}_k}{\Omega^{D-1}}
 \sum_{\rm pol} \epsilon_c^{\alpha} \epsilon_c^{\beta *}= 
  \frac{D-2}{D-1} \Pi_{\alpha\beta}(P)                    \; ,
\ee                                                           
one can easily compute the sum over {\em effective} polarizations   
\be
\int \frac{d\Omega^{D-1}_k}{\Omega^{D-1}}
    \sum_{\epsilon_c}  \epsilon^{J}_{{\rm eff}, \alpha} 
                       \epsilon^{J *}_{{\rm eff}, \beta}
        = -  N_J \frac{D-2}{D-1} \Pi_{\alpha\beta}(P) \; ,
\ee
where $N_J$ is the number of degrees of freedom of the $^3P_J$ state in $D$
dimensions. This shows that 
the last term in eq.~(\ref{eq:dAdq}), once squared and averaged
on the directions of the outgoing gluon, is proportional to the
amplitude of a $^3S_1$ state coupled to two gluons , which vanishes by 
$C$-parity.  As a result we obtain the factorized expression:
\be                                                           
    \sum_{\rm col, pol} \; \vert\overline{\aeights}\vert^2 \=
     \ca g^2 \left[\frac{2\sp{a}{P} }{(\sp{a}{k} )  (\sp{P}{k} )} -
      \frac{M^2}{(\sp{P}{k} )^2 } \right]\sum_{\rm col, pol} \;
\vert\overline{\aeightb}\vert^2 
\label{eq:cofact}
\ee                                              
The amplitudes for $\gamma q \to \q q$ are IR finite, and there is no need to
                study their soft behaviour for our applications.

\subsection{Kinematics and factorization of soft and collinear singularities}
The kinematics of the process $k + p_1 \to P + p_2 $, 
where $P$ is the momentum of the heavy quark pair, $k$ the momentum of
the photon and  $p_i$ are the momenta of the massless
partons, can be described in terms of the standard Mandelstam variables
$s$, $t$ and $u$:
\ba
     s &=& (k  + p_1)^2 \; , \\      
     t &=& (p_1- p_2)^2 \;\equiv\; -\frac{s}{2}(1-x)(1-y) \; , \\
     u &=& (k  - p_2)^2 \;\equiv\; -\frac{s}{2}(1-x)(1+y) \; .   
\ea                                                          
Here we introduced the Lorentz-invariant dimensionless variables
$x=M^2/s$ and $y$ ($-1<y<1$), defined by the above equations.
In the center-of-mass frame of the partonic collisions, the variable $y$
becomes the cosine of the scattering angle $\theta$. In terms of $x$ and $y$
the total partonic cross-section can be written in $D$ dimensions as follows:
\be                
   \sigma \= \frac{1}{2s} \int d\Phi_{(2)}(x,y) \; \m(x,y) \; ,
\ee                                                        
where $\m=\overline{\sum}\vert A\vert^2$ is the spin- and colour-averaged matrix
element squared in $D$ dimensions and $d\Phi_{(2)}(x,y)$ is the $D$-dimensional
two-body phase space:                                           
\be
  d\Phi_{(2)}(x,y) \= \frac{4^{\eps}}{K} \, 
      \left(\frac{4\pi}{s}\right)^{\eps} \, \Gamma(1+\eps) \,
      \frac{1}{16\pi} \, (1-x)^{1-2\eps} \, (1-y^2)^{-\eps} \, dy \; ,
\ee                                                                   
with
\be
  K=\Gamma(1+\epsilon)\, \Gamma(1-\epsilon) = 1+\epsilon^2\frac{\pi^2}{6}+
  {\cal O}(\epsilon^3)\; .
\ee
The soft and collinear singularities are associated to the vanishing of $t$ or
$u$, which appear at most as single poles in the expression of \m. One can
therefore introduce the finite, rescaled amplitude squared \mbar:
\be
     \m \= \frac{1}{ut} \, \mbar \= \frac{4}{s^2(1-x)^2(1-y^2)} \, \mbar \; .
\ee                                                                
In terms of \mbar, the partonic cross-sections read as follows:
\ba                                                 
  &&   \sigma(x) \=
     \frac{4C}{s^2} \, (1-x)^{-1-2\eps} \, \int_{-1}^{1} \, dy \,
     (1-y^2)^{-1-\eps} \, \mbar(x,y)\, dy \; , \\
  && C \= \frac{4^{\eps}}{K} \,  
      \left(\frac{4\pi}{s}\right)^{\eps} \, \Gamma(1+\eps) \,
      \frac{1}{32\pi s}  \; .               
\ea                          
Soft and collinear singularities are now all
contained in the universal poles which develop as $x\to 1$ and $y^2\to 1$. The
residues of these poles can be derived without an explicit calculation of the
matrix elements, as they only depend on the  universal structure of collinear
and soft singularities. We will carry out an explicit evaluation of these
residues in the next section.
                               
\section{Results}
\label{sec:production}
\subsection{$g\gamma \to g\o^{[8]}$ processes}

\begin{figure}[t]
\begin{center}
\epsfig{file=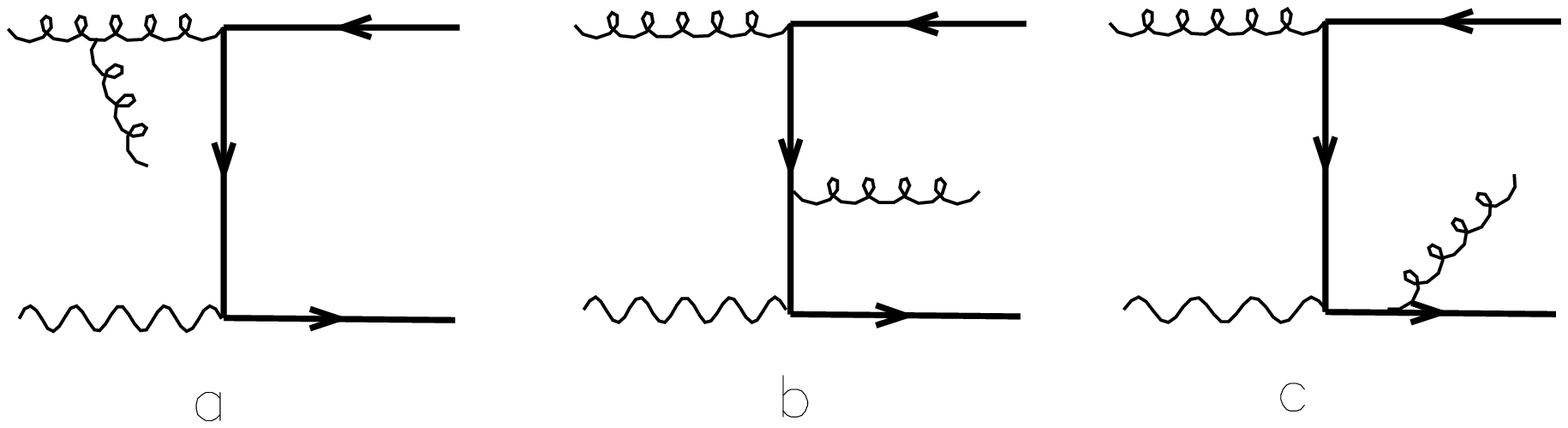,width=0.75\textwidth,clip=}
\ccaption{}{\label{fig:gph-g} Diagrams for the real corrections to the 
$g\gamma$ channels. Permutations of outgoing gluons 
and/or reversal of fermion lines are always implied.}
\end{center}
\end{figure}

We start by considering the soft limit, $x\to 1$. 
 The
following distributional identity holds for small $\eps$:
\be                                       
     (1-x)^{-1-2\eps} \= -\frac{\beta^{-4\eps}}{2\eps} \delta(1-x) +
        \left( \frac{1}{1-x} \right)_\rho \, - \,
    2\eps \left( \frac{\log(1-x)}{1-x}\right)_\rho \, + \, {\cal{O}}(\eps^2) \, 
\ee                                                                           
where $\rho=M^2/S_{\gamma h}$, $h$ being the initial state target hadron, and
$\beta=\sqrt{1-\rho}$. The $\rho$-distributions are defined by:          
\be                                                                       
   \int_{\rho}^{1} \, dx \; \left[d(x)\right]_{\rho} t(x) \=
   \int_{\rho}^{1} \, dx \; d(x) \; \left[t(x)-t(1)\right] \; .
\ee                                                   
We can therefore write, with obvious notation,
\be
   \sigma(x) \= \sigma_{x=1} + \sigma_{x\neq 1} \; .
\ee                                                
The first term on the right-hand side is given by the following expression:
\be  \label{eq:sigx1}
  \sigma_{x=1} \=    
     -\frac{4C}{M^4} \, \frac{\tau^{-4\eps}}{2\eps} \, \delta(1-x) \,
      \int_{-1}^{1} \, dy \,               
     (1-y^2)^{-1-\eps} \, \mbar(x=1,y)\; .
\ee                                       
The $x\to 1$ limit of \mbar\ can be easily derived from eq.~(\ref{eq:cofact}):
\ba                                                          
  \mbar(x,y) & \stackrel{x\to 1}{\longrightarrow} & 
       s \, g^2 \, \ca \,(1-y)^2  \mborn \; ,
\ea                                                                    
where \mborn\ is the $D$-dimensional Born amplitude squared for the \
$\gamma g\to \o^{[8]}$
process, which is independent of $y$. The integration over $y$ of
eq.~(\ref{eq:sigx1}) is elementary, and leads to the following result:
\be                 
   \sigma_{x=1} \=                          
      \left(\frac{4\pi\mu^2}{s}\right)^{\eps} 
      \frac{\Gamma(1+\eps)\, \tau^{-4\eps}}{2\eps^2} \,
      H \, \ca \, \frac{\as}{\pi} \, \sborn \;,
\ee                                  
where $\as$ is related to the $D$-dimensional bare coupling $\asb$ and to the
renormalization scale $\mu$ by $\as\mu^{2\eps}=\asb =g^2/4\pi$.
$H$ is defined by               
\be              
   H \= \frac{\Gamma(2-\eps)}{\Gamma(1+\eps)\Gamma(2-2\eps)} \=
      1 + \epsilon + 2 \epsilon^2  
     - \frac{\pi^2}{3} \eps^2 + {\cal O}(\eps^3) \; ,
\ee                                                                  
and \sborn\ is the $D$-dimensional Born cross-section:
\be 
  \sborn \= \pi/M^4 \mborn
\delta(1-x) \; \equiv \; \sborno \delta(1-x) \; .
\label{eq:sigmaB}
\ee
The collinear singularities remaining in $\sigma_{x\neq 1}$ can be factored out
by using the following distributional identity:            
\be
   (1-y^2)^{-1-\eps} \= -\left[ \delta(1-y)+\delta(1+y)\right]
  \frac{4^{-\eps}}{2\eps} 
  + \frac{1}{2}\left[\left(\frac{1}{1-y}\right)_+ + 
    \left(\frac{1}{1+y}\right)_+ \right] + {\cal O}(\eps) \; ,
\ee                                                           
where the distributions on the right-hand side are defined by:
\be                            
   \int_{-1}^{1} \, dy \, \left(\frac{1}{1\pm y}\right)_+ t(y) \=
   \int_{-1}^{1} \, dy \, \frac{1}{1\pm y} \left[t(y)-t(\mp 1)\right] \; .
\ee 
The  contribution $\sigma_{x\neq 1}$ can then be split into two terms:
\be                                              
   \sigma_{x\neq 1} \= \sigma_{y=1}+\sigma_{\rm finite} \; .
\ee                                                                      
The term $\sigma_{\rm finite}$ has no residual divergences, and is given by the
following expression:                 
\be  \label{eq:finite}
  \sigma_{\rm finite} \= \frac{2C}{s^2} \left( \frac{1}{1-x}\right)_{\rho}
  \, \int_{-1}^{1} \, dy \, \left[\left(\frac{1}{1-y}\right)_+ +       
    \left(\frac{1}{1+y}\right)\right] \,  \, \mbar(x,y) \; .        
\ee                                                         
We removed the $+$ from the $1/(1+y)_+$ distribution because no collinear
singularity can arise from the emission of the gluon collinear to the photon, 
and $\mbar(x,-1)=0$.
$\sigma_{\rm finite}$ explicitly depends on the nature of the quarkonium state
produced.            
For processes whose Born contribution vanishes, this is the only non zero term.
$\sigma_{y=1}$ is given by:
\be             
  \sigma_{y=1} \=
  - \frac{4C}{s^2} \, \frac{4^{-\eps}}{2\eps} \,
  \left[ \left( \frac{1}{1-x} \right)_+        
       - 2\eps \left( \frac{\log(1-x)}{1-x}\right)_+ \right] 
     \, \mbar(x,y=1) \; .
\ee               
The limit for $y\to 1$ of $\mbar(x,y)$ is universal, thanks to the
factorization of collinear singularities:
\be
  \mbar(x,y) \stackrel{y\to 1}{\longrightarrow} 8\pi \, s \, \asb P_{gg}(x)
  \frac{1-x}{x} \mborn \; .                                       
\ee
Using this relation we get:
\be   \label{eq:sigcol} 
  \sigma_{y=1} \=
  - \frac{1}{\epsbar} \, \left(\frac{\mu^2}{s}\right)^{\eps} \,
    \frac{\as}{2\pi} \, \pgg{x}  \, (1-x)x \,  \sborno 
    \, \left[ \left( \frac{1}{1-x} \right)_+        
       - 2\eps \left( \frac{\log(1-x)}{1-x}\right)_+ \right] \; ,
\ee 
with
\be
P_{gg}(x) =
2\ca\left[\frac{x}{1-x}+\frac{1-x}{x}+x(1-x)\right]\, .
\ee
and 
\be
   \frac{1}{\epsbar} \= \frac{1}{\eps} -\gamma_{\rm E} + \log(4\pi) \; ,
\ee
The collinear poles take the form dictated by the factorization theorem.
According to this the partonic cross-section can be written as:
\ba  \label{eq:isfact}         
     {\rm d}\sigma_{\gamma j}(p_\gamma,p_h) &=& \sum_l
     {\rm d}\hat\sigma_{\gamma l}(p_\gamma,x p_h)     
     \Gamma_{lj}(x) {\rm d}x \;,                
\\                    
     \Gamma_{ij}(x)  &=&
      \delta_{ij} \delta(1-x) \;-\; \frac{1}{\epsbar} \,\frac{\as}{2\pi} 
      \left(\frac{\mu^2}{\mufsq}\right)^{\epsilon}
      {\cal P}_{ij}(x) \; + \; K_{ij}(x) \; ,
\ea                                   
where ${\rm d}\hat\sigma$ is free of collinear singularities as $\epsilon\to
0$. Here we allowed the factorization scale $\muf$ to differ from the
renormalization scale $\mu$.
The functions ${\cal P}_{ij}(x)$ are the $D=4$  Altarelli-Parisi splitting
kernels, collected in Appendix~\ref{appA}, and the factors $K_{ij}$ are
arbitrary functions, defining the factorization scheme. In this paper we adopt
the \MSB\ factorization, in which $K_{ij}(x)=0$ for all $i,j$. For the
definition of $K_{ij}(x)$ in the DIS scheme, see for example
ref.~\cite{Kuehn93}. 

\begin{figure}[t]
\begin{center}
\epsfig{file=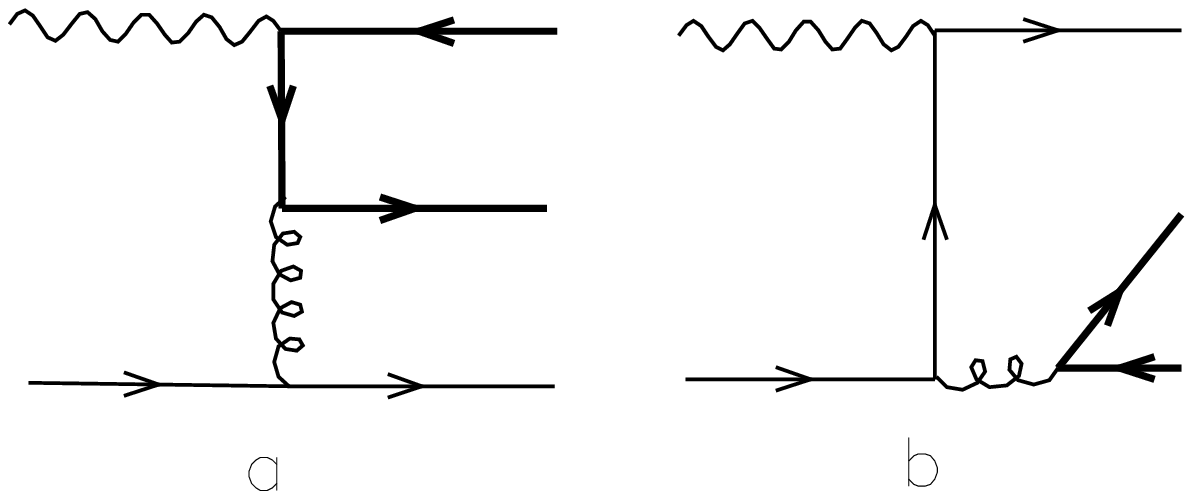,width=0.75\textwidth}
\ccaption{}{\label{fig:gq-q} Diagrams for  the $\gamma q$
channels. Reversal of fermion lines is always implied.}
\end{center}
\end{figure}

Expanding eq.~(\ref{eq:isfact}) order-by-order in $\as$, we extract the
counter-term $\sigma^{(c)}_{y=1}$, defined by:
\be \label{eq:sigct}
  \sigma^{(c)}_{y=1} \= \frac{1}{\epsbar} \,\frac{\as}{2\pi} \left(
  \frac{\mu^2}{\mufsq}\right)^{\eps} \, \cpgg{x}\, x \sborno \; ,
\ee                                                  
with
\be
{\cal P}_{gg}(x) =
2\ca\left[\frac{x}{(1-x)_{\rho}}+\frac{1-x}{x}+x(1-x)\right]+\left(\b0+4
\ca\log\beta\right)\delta(1-x)\, .
\ee
Putting all pieces together, we come to the final result for the real-emission
cross-section:                                                        
\ba
\sigma^H[g\gamma \to \q^{[8]}g ](x) &=&
       \frac{\as}{2\pi} \sigma_0^H[g\gamma \to \q^{[8]}] \nn \\
&\times& \left\{  
f_\epsilon(s)\left[\ca\left(\frac{1}{\ep^2}+\frac{17}{6\ep}  
+2 -\frac{\pi^2}{3}-4 \log \tau + 8 \log^2\tau\right) -\frac{2}{3\ep}\nf\tf
\right]\delta(1-x)+\right.\nn\\ 
&+&\left.
\left[x {\cal P}_{gg}(x)\log\frac{s}{\mufsq} + 
  2 x(1-x) P_{gg}(x)\left(\frac{\log(1-x)}{1-x}\right)_{\rho} +
   \left(\frac{1}{1-x}\right)_{\rho} f_{\gamma g}[\q^{[8]}](x)\right]\right\}
\nn\\                                                     
\nn\\
&& [\q^{[8]}=\etah,\chizh,\chith]  \; ,        
\ea
where $f_\eps(s)$ is defined in Appendix~\ref{appA} and
$\sigma_0^H[\gamma g \to \q^{[8]}]$ 
is the $D$-dimensional, Born-level partonic
cross-section for the production of the quarkonium state $H$ via the
$\q^{[8]}$ intermediate state, after removal of the $\delta(1-x)$ term (see
eq.~(\ref{eq:sigmaB})).        
The finite functions $f_{\gamma g}(x)$, obtained from the explicit evaluation of
eq.~(\ref{eq:finite}), are collected in Appendix~\ref{appNLO}.

\subsection{$q \gamma \to q\q^{[8]}$ processes}                              
The Born-level processes $q\gamma \to \q^{[8]}$ identically vanish. 
As a result IR divergences at $O(\assq \aem)$ and virtual 
corrections are not present. The only possible singularities appearing at this
order come from the emission of the 
final-state quark collinear to the initial-state one or to the photon..
The behaviour of the amplitude (fig.~\ref{fig:gq-q}(a))               
for $\q^{[8]}=\etah,\chizh,\chith $ 
in the $y\to 1$ collinear limit is again
controlled  by the Altarelli-Parisi splitting functions:
\be                                                     
  \mbar(x,y) \stackrel{y\to 1}{\longrightarrow} 8\pi \, s \, \asb 
  P_{gq}(x) \frac{1-x}{x} \mborn \; .
\ee                                  
In analogy to the $\gamma g$ case, one introduces 
the following counter-term in the 
\MSB\ scheme:                                             
\be \label{eq:sigctqq}
  \sigma^{(c)}_{y=1} \= \frac{1}{\epsbar} \,\frac{\as}{2\pi} \left(
  \frac{\mu^2}{\mufsq }\right)^{\eps} \, \cpgq{x}\, x\sborno \; ,
\ee                                                         
where $\cpgq{x}\ $ is defined in Appendix~\ref{appA}.
                                       
Following a procedure analogous to the one detailed in the case of $\gamma g$
production, we find the following result:  
\ba
\sigma^H[\gamma\,q\to \q^{[8]} \,q](x) &=& \frac{\as}{\pi}
\sigma_0^H[\gamma g\to \q^{[8]}]\times \nn\\  
&&\left\{\left[\frac{x}{2}             
   P_{gq}(x)\log\frac{s(1-x)^2}{\qf} + \cf\frac{x^2}{2}\right]+
   f_{\gamma q}[\q^{[8]}](x)\right\}\, , \nn\\
&& [\q^{[8]}=\etah,\chizh,\chith] \label{eq:qgP1}\; ,
\ea                                              
where the functions $f_{\gamma q}(x)$ are collected in
Appendix~\ref{appNLO}, together with the result for 
$\threePone$
production, for which no collinear singularity is present to start with.

\begin{figure}
\begin{center}
\epsfig{figure=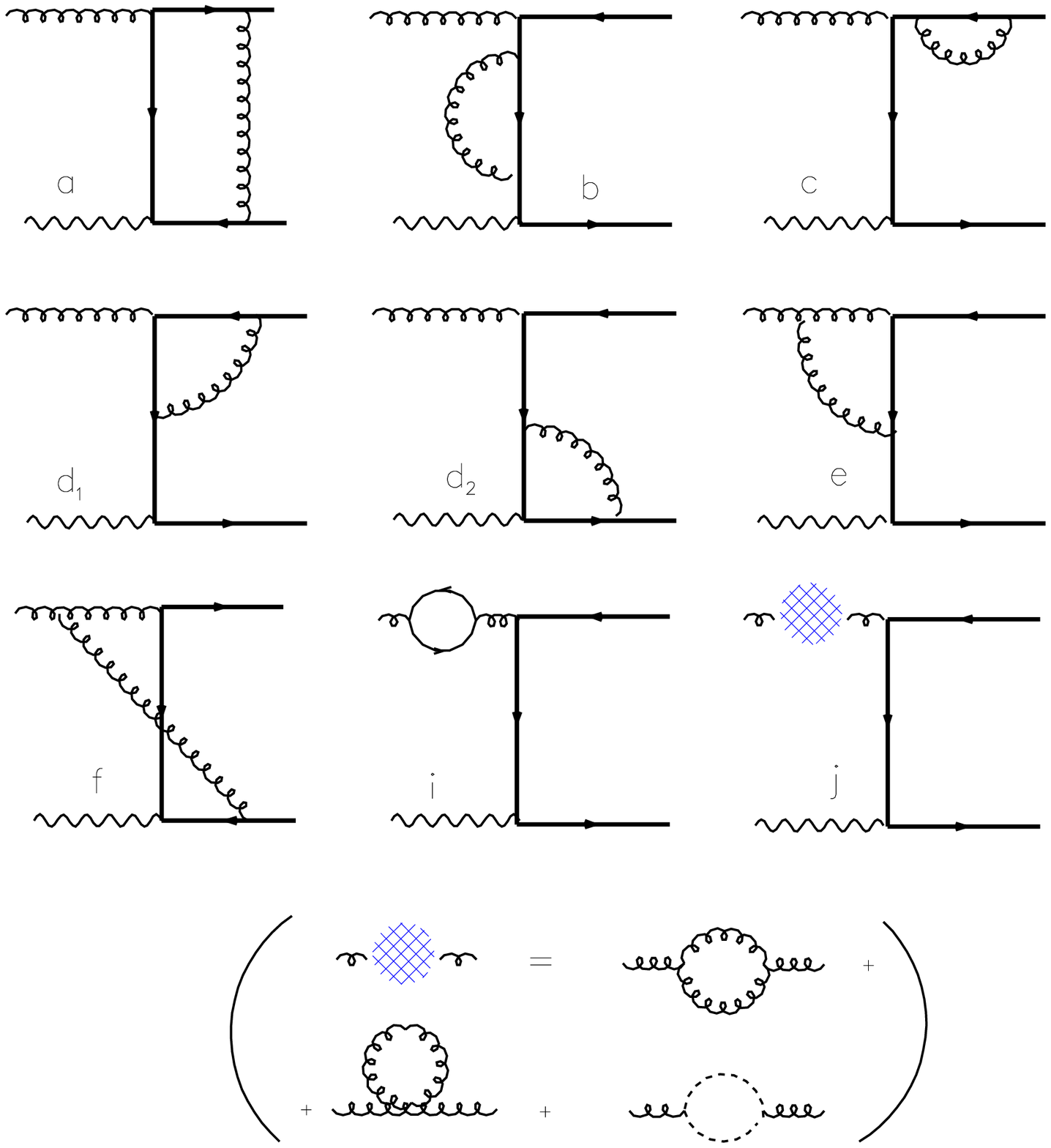,width=0.75\textwidth}
\ccaption{}{ \label{fig:gg}
Feynman diagrams contributing to the one-loop 
corrections to the processes $g \gamma \to\QQ[^3P_0^{[8]}] $, 
$g \gamma \to\QQ[^3P_2^{[8]}]$ and $ g \gamma \to\QQ[^1S_0^{[8]}]$. }
\end{center}
\end{figure}

For $\psih$ production only the diagram in fig.~\ref{fig:gq-q}(b) gives
a non vanishing contribution.                  
In the $y\to 1$ collinear limit we have:
\be                                                     
  \mbar(x,y) \stackrel{y\to 1}{\longrightarrow} 8\pi \, s \, \aemb e_Q^2
  P_{q\gamma}(x) \frac{1-x}{x} \mborn \; .                              
\ee                                  
where $\mborn$ is the Born amplitude for the process $\qq \to \psih$
and $P_{q\gamma}(x)$ is the Altarelli-Parisi photon-splitting function in $D$
dimensions. Introducing the following counter-term in the 
\MSB\ scheme:                                             
\be \label{eq:sigctqph}
  \sigma^{(c)}_{y=1} \= \frac{1}{\epsbar} \,\frac{\aem e_Q^2}{2\pi} \left(
  \frac{\mu^2}{\mug^2}\right)^{\eps} \, \cpqgamma{x}\, x\sborno \; ,
\ee                                                         
where $\cpqgamma{x} \ $ is defined in Appendix~\ref{appA},
we can write the following partonic cross-section:
\ba                                     
\dsh[\, \gamma \,q\to \psih\,q](x)&=& \frac{\aem \ech}{\pi}\szh[\qq\to\psih] \nn\\
&& \times\left\{\left[\frac{x}{2} 
P_{q\gamma}(x)\log\frac{s(1-x)^2}{\mu_{\gamma}^2} + \df
x^2(1-x)\right]+ f_{\gamma q}[\psih](x)\right\}
\ea                                                               
Having absorbed the collinear $\gamma\to q$ divergence in the photon
structure function, the total cross-section will now include a piece
proportional to the resolved component of the photon (see, \eg,
ref.~\cite{Frixione94,Frixione97}):
\ba                                                 
     {\rm d}\sigma_{\gamma \, h}(p_1,p_h) &=& 
     \sum_{i}                           
     {\rm d}\hat\sigma_{\gamma i}(xS,\muf,\mug) F_{i\,h}(x,\muf) {\rm d}x \nn\\
&&+  \sum_{k,l}                                        
     {\rm d}\hat\sigma_{kl}(x_1 p_\gamma,x_2 p_h,\muf)  
     F_{k\gamma}(x_1,\mug) F_{lh}(x_2,\muf) {\rm d}x_1 {\rm d}x_2 \;,
\ea                             
where $F_{k\gamma}(x,\mug)$ 
is the density of the parton $k$ in the photon, and
$F_{kh}(x,\muf)$ is the density of the parton $k$ in the hadron $h$.
The full $\oacube$ expression for the parton-parton $\psih$ cross-sections
to be used in the resolved-photon contribution to the above equation can be
found in~\cite{Petrelli97}.

{\renewcommand{\arraystretch}{1.5}
\begin{table}
\begin{center} 
\begin{tabular}{|l|l|l|} \hline
Diag. &${\cal D}_k$ &$f_k^{(8)}$\\ \hline\hline

 a& $\frac{\pi^2}{2v} +\frac{1}{\ep} -2 +2\log 2  $
&$\caf$\\ \hline
 b& $  -\frac{1}{2 \eu}-1 + 3\log 2 $&\cf  \\ \hline
 c&  $  -\frac{1}{2 \eu}-\frac{1}{\ep}-2-3\log 2 $ &\cf\\ \hline
 ${\rm d}_1$ & $  \frac{1}{2 \eu} - \log 2 +\frac{\pi^2}{8} $ &\cf\\ \hline
 ${\rm d}_2$ & $  \frac{1}{2 \eu} - \log 2 +\frac{\pi^2}{8} $ &\caf\\ \hline 
 e & $ +\frac{3}{2\eu}-\frac{1}{2\ep^2}-\frac{1}{2\ep}+2- \log 2 +\frac{\pi^2}{12} $ &$\frac{1}{2}$\ca\\ \hline 
 f & $  -\frac{1}{2\ep^2}-\frac{1}{2\ep}
 -1+2\log 2+\frac{5}{24}\pi^2 $&$\frac{1}{2}\ca $ \\\hline
 i & $ \frac{5}{12\eu}-\frac{5}{12\ep}  $ &\ca \\\hline
 j & $ \left( -\frac{1}{3\eu}+\frac{1}{3\ep}\right) \nf $ &\tf \\ \hline
\end{tabular}
\ccaption{}{\label{tab:1s0} ``Diagrammatic''  
partial virtual QCD corrections to the processes 
$ g \gamma \to [^1S_0^{[8]}]$ 
(diagram multiplicities are included)}
\end{center}
\end{table}

{\renewcommand{\arraystretch}{1.5}
\begin{table}
\begin{center} 
\begin{tabular}{|l|l|l|} \hline
Diag. &${\cal D}_k$ &$f_k^{(8)}$\\ \hline\hline

 a & $ \frac{\pi^2}{2v} +\frac{1}{\ep} -\frac{4}{9} +\frac{32}{9}\log 2
    +\frac{\pi^2}{12}  $&$\caf$\\ \hline
 b & $  -\frac{1}{2 \eu}-\frac{13}{9} +\frac{5}{9}\log 2 $&\cf \\ \hline
 c & $  -\frac{1}{2 \eu}-\frac{1}{\ep}-2-3\log 2 $ &\cf\\ \hline
 ${\rm d}_1$ & $  \frac{1}{ 2\eu} +\frac{7}{9}-\frac{5}{9}\log 2 +
   \frac{\pi^2}{12} $&\cf \\ \hline
 ${\rm d}_2$ & $  \frac{1}{ 2\eu} +\frac{7}{9}-\frac{5}{9}\log 2 +
   \frac{\pi^2}{12} $ &\caf \\ \hline
 e & $ \frac{3}{2\eu}-\frac{1}{3\ep^2}-\frac{10}{9\ep}+ \frac{11}{6}-
   \frac{1}{3}\log 2 +\frac{\pi^2}{18} $ &$\frac{1}{2}$\ca\\ \hline 
 f & $  -\frac{2}{3 \ep^2}+\frac{1}{9\ep}-\frac{1}{2}+
   \frac{10}{3}\log 2+\frac{5}{18}\pi^2 $&$\frac{1}{2}$\ca\\ \hline
 i & $ \frac{5}{12\eu}-\frac{5}{12\ep}  $ & $\ca$\\\hline
 j & $ \left( -\frac{1}{3\eu}+\frac{1}{3\ep}\right) \nf $ &\tf\\ \hline
\end{tabular}
\ccaption{}{\label{tab:3p0} ``Diagrammatic'' partial virtual QCD corrections 
to the processes $g\gamma~\to~[^3P_0^{[8]}]$ 
(diagram multiplicities are included).}
\end{center}
\end{table}

{\renewcommand{\arraystretch}{1.5}
\begin{table}
\begin{center} 
\begin{tabular}{|l|l|l|} \hline
Diag. &${\cal D}_k$ &$f_k^{(8)}$\\ \hline\hline

 a & $ \frac{\pi^2}{2v} +\frac{1}{\ep} -\frac{5}{3} +\frac{7}{3}\log 2
    +\frac{\pi^2}{8}  $&$\caf$\\ \hline
 b & $  -\frac{1}{2 \eu}-\frac{1}{6} +\frac{11}{6}\log 2 $&\cf \\ \hline
 c & $  -\frac{1}{2 \eu}-\frac{1}{\ep}-2-3\log 2 $ &\cf\\ \hline
 ${\rm d}_1$& $  \frac{1}{2\eu} -\frac{1}{12}-\frac{7}{12}\log 2 
  -\frac{\pi^2}{16} $&\cf \\ \hline
 ${\rm d}_2$& $  \frac{1}{2\eu} -\frac{1}{12}-\frac{7}{12}\log 2 
  -\frac{\pi^2}{16} $ &\caf \\ \hline
 e & $ \frac{3}{2\eu}-\frac{3}{16\ep^2}-\frac{17}{32\ep}+
     \frac{59}{192}+\frac{\pi^2}{32} $ &$\frac{1}{2}$\ca\\ \hline 
 f & $  -\frac{13}{16 \ep^2}-\frac{15}{32\ep}
    -\frac{107}{192}+\frac{11}{4}\log 2+\frac{67}{96}\pi^2 $&$\frac{1}{2}\ca$
    \\\hline
 i & $ \frac{5}{12\eu}-\frac{5}{12\ep}  $ &$\ca$\\ \hline
 j & $  \left(-\frac{1}{3\eu}+\frac{1}{3\ep}\right) \nf $ &\tf\\ \hline
\end{tabular}
\ccaption{}{\label{tab:3p2}
``Diagrammatic'' partial virtual QCD corrections 
to the processes $g\gamma\to [^3P_2^{[8]}]$ 
(diagram multiplicities are included).}
\end{center}
\end{table}

\subsection{Virtual corrections}
We present in this section the results of the calculation of the 1-loop
diagrams necessary for the evaluation of                            
the virtual corrections to the production matrix elements.
These results can be obtained in a straightforward way from the calculation of
the virtual corrections to 1-loop hadroproduction, presented in
ref.~\cite{Petrelli97}. We shall therefore limit ourselves to presenting the
final answers. The relevant Feynman diagrams
are shown  in figures ~\ref{fig:gg}, 
and the results are given diagram by diagram
in tables~\ref{tab:1s0}, \ref{tab:3p0}, 
\ref{tab:3p2}. 
In these tables we report the contribution of each diagram $k$,
indicating separately the colour factors \fk.
The expressions \dk\ appearing in the tables are defined by the following
equation:
\be
\sigma_V^H[ij\to\q] = \frac{\as\mu^{2\ep}}{\pi}\sigma_0^H[ij\to\q]
\,  f_{\epsilon}(s) \sum_k {\cal D}_k f_k\,\delta(1-x). 
\ee
where the sum extends over the set of diagrams and $f_\eps(s)$ is defined in
Appendix~\ref{appA}.

 The singularity structure of the virtual corrections is dictated by the
renormalization properties of the theory, by the universal form of the Coulomb
limit, and by requirement that soft and collinear singularities cancel against
the real corrections evaluated above. The form of the virtual corrections to
the cross-section is therefore the following:
\be      
         \sigma^{(V)} \= \sigma_0
         \frac{\as}{2\pi}           
     f_\epsilon(s)
\times\label{eq:virtual}                              
    \left\{ \frac{b_0}{\epsuv} + (\caf)\frac{\pi^2}{v} -
     \ca \left( \frac{1}{\epsir^2} + \frac{17}{6\epsir} \right)
      +  \frac{2}{3\epsir} \nf \tf + 2\,D^{[8]}_{\o}\right\}\; ,
\ee   
where we explicitly labelled the $\eps$'s to indicate their origin,
and where all of the state dependence is included in the finite factor 
$D_{\q}$. The quark-antiquark relative velocity $2v$ and $b_0$ 
are                                 
defined in Appendix~\ref{appA}. $\nf$ is the number of flavours lighter than
the heavy, bound one.
 
Summing the contribution of all diagrams, we obtain the following results for
the colour-octet coefficients $D_{\q}^{[8]}$ :   
\ba                             
     && D^{[8]}_{\oneSzero} \= \; \cf\left(-5+\frac{\pi^2}{4}\right)+
        \ca\left(\frac{3}{2}+\frac{\pi^2}{12}\right)  \\
     && D^{[8]}_{\threePzero}          
             \=  \cf \left( -\frac{7}{3}+\frac{\pi^2}{4} \right) +
        \ca \left( \frac{1}{2}+\frac{\pi^2}{12}\right) \\
     && D^{[8]}_{\threePtwo} \=  -4\cf +
                          \ca \left( \frac{3}{4}+\frac{\log 2}{2} +
                          \frac{\pi^2}{3} \right) \; .
\ea
The final results for the finite sums of real plus virtual corrections are
collected in Appendix~\ref{appNLO}.


\section{Phenomenology}
\label{sec:pheno}
In this section we study some of the properties of the higher-order corrections
calculated in this paper, and their effect on typical Born-level predictions. A
more thorough phenomenological study including a comparison to currently
available data and fits to the non-perturbative parameters will be presented
elsewhere. 

{
\begin{figure}
\begin{center}
\epsfig{figure=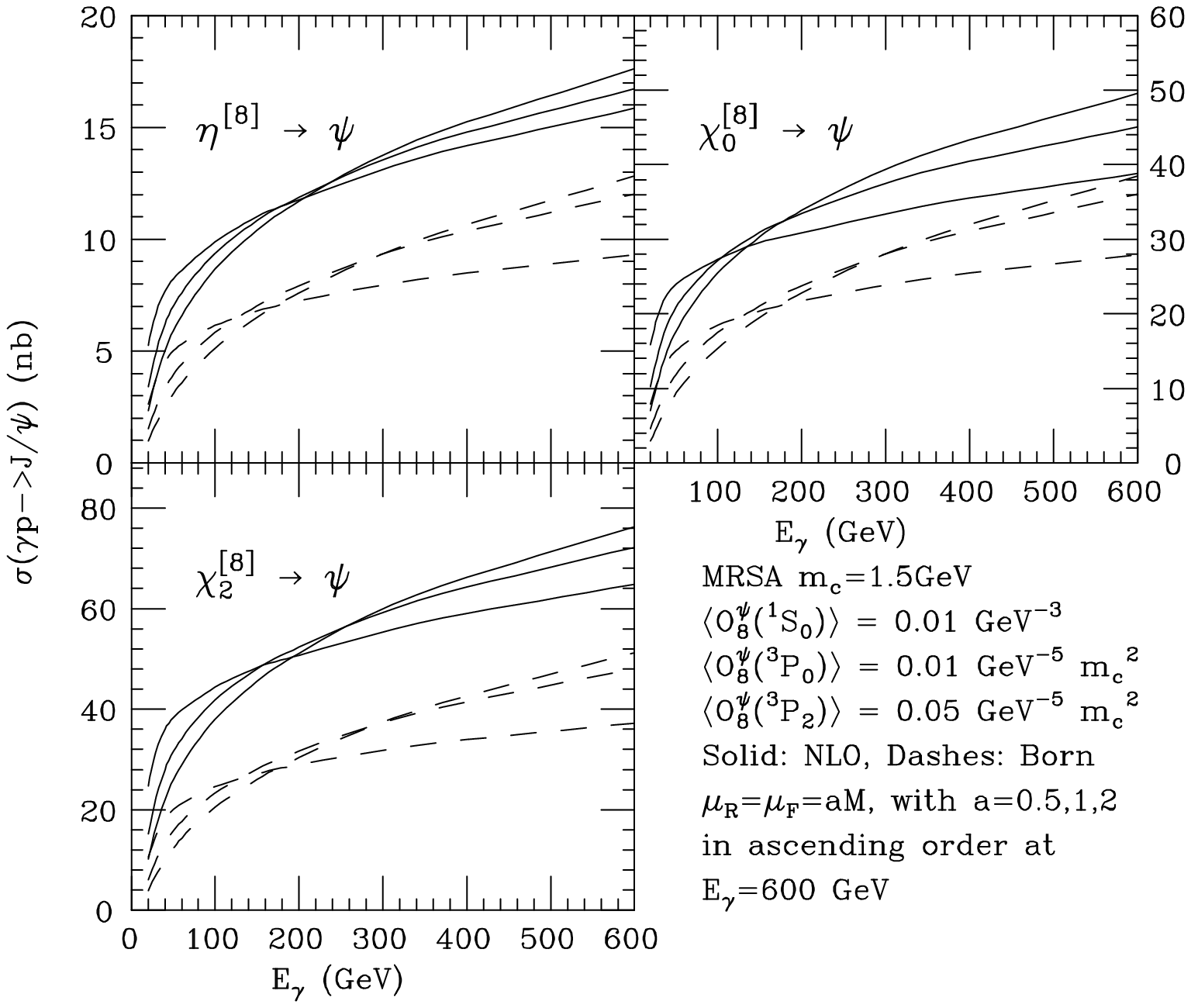,width=0.7\textwidth}
\ccaption{}{ \label{fig:totfxt}
Born and
NLO total photoproduction cross-sections as a function of the photon beam
energy.}
\end{center}
\end{figure}

\nopagebreak
\begin{figure}
\begin{center}
\epsfig{figure=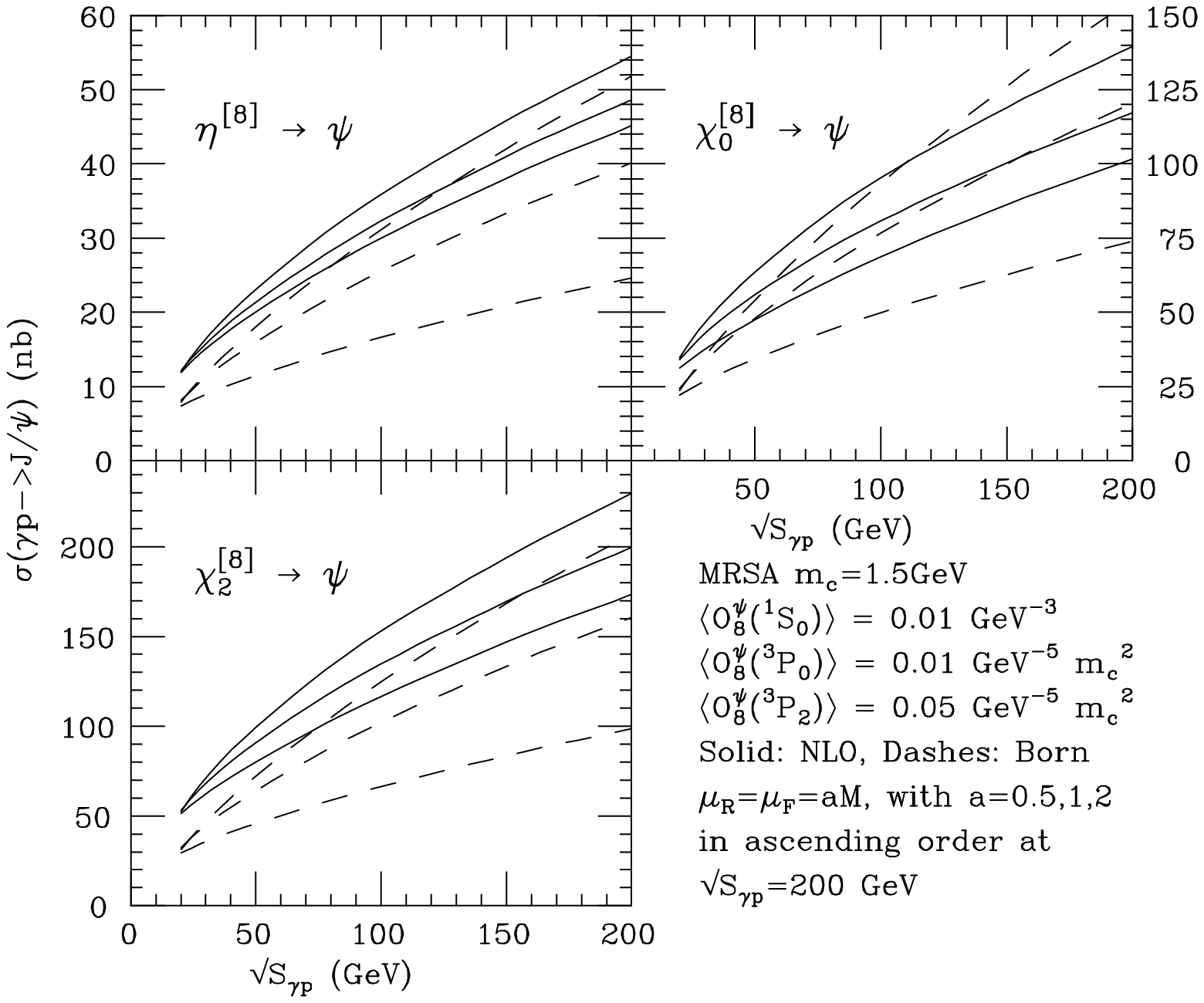,width=0.7\textwidth}
\ccaption{}{ \label{fig:totcol}    
Born and                      
NLO total photoproduction cross-sections as a function of the $\gamma$p CM
energy.}
\end{center}
\end{figure}  }
               
To start with, we show in fig.~\ref{fig:totfxt} the comparison between Born and
NLO total photoproduction cross-sections as a function of the photon beam
energy, in the energy range of fixed-target experiments. 
Here and in the following, we concentrate on the colour-octet contributions of
$\oneSzero$, $\threePzero$ and $\threePtwo$ states, which are by far the
dominant processes.
We fix   $m_c=1.5$~GeV and we take, for the sake of
definiteness~\cite{Cacciari96},             
$\opetah=0.01$~GeV$^{-3}$ and $\opchizh/m_c^2=0.01$~GeV$^{-5}$, with
$\opchijh=(2J+1)\opchizh$.
                          
\begin{figure}
\begin{center}
\epsfig{figure=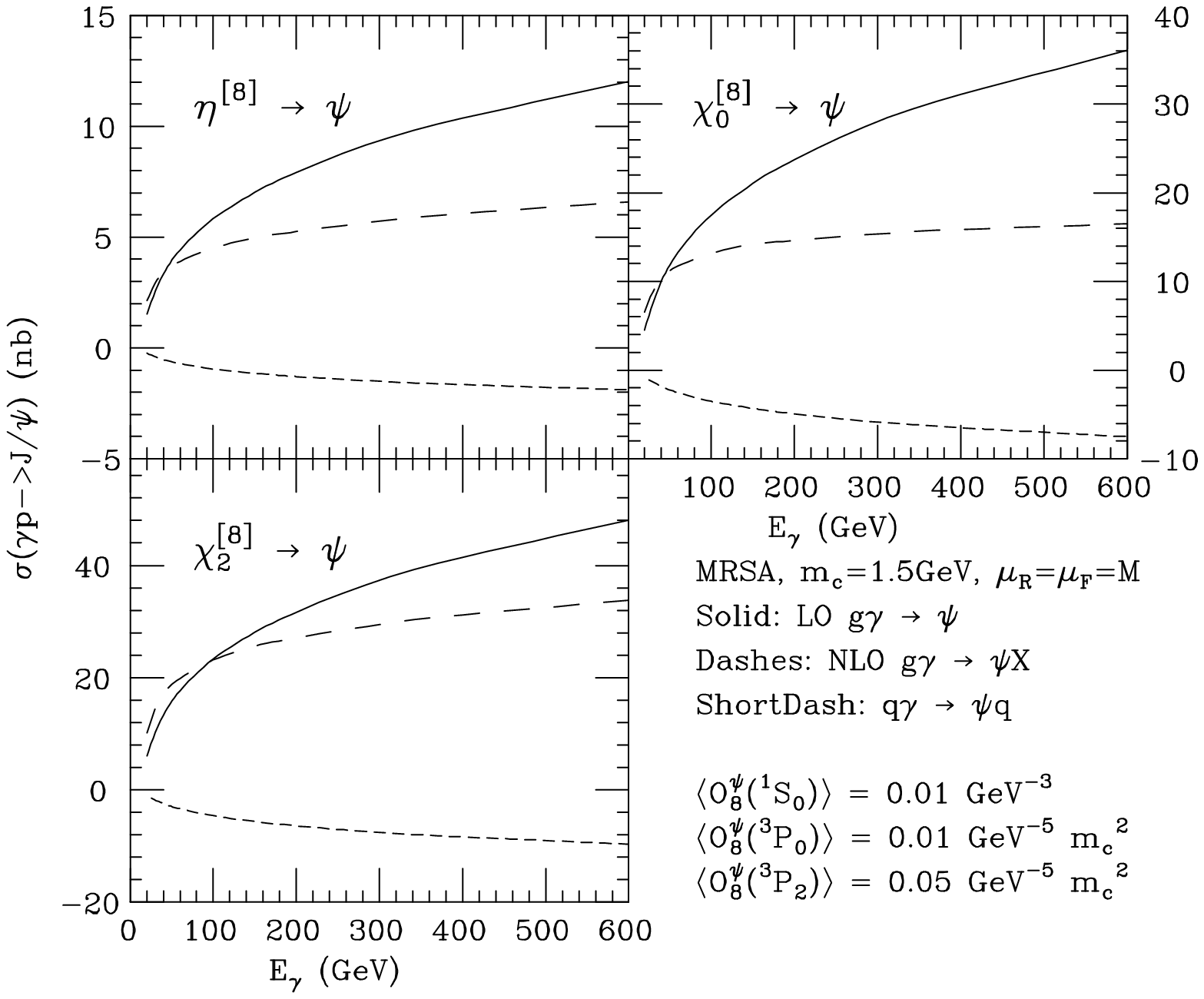,width=0.70\textwidth}
\ccaption{}{ \label{fig:qgfxt}     
Separate contributions from Born, $\oatwoa$ $\gamma g$ and $\oatwoa$ $\gamma q$ 
processes to
total photoproduction cross-sections as a function of the photon beam energy.}
\end{center}                                                                  
\end{figure}

{ \nopagebreak
\begin{figure}
\begin{center}
\epsfig{figure=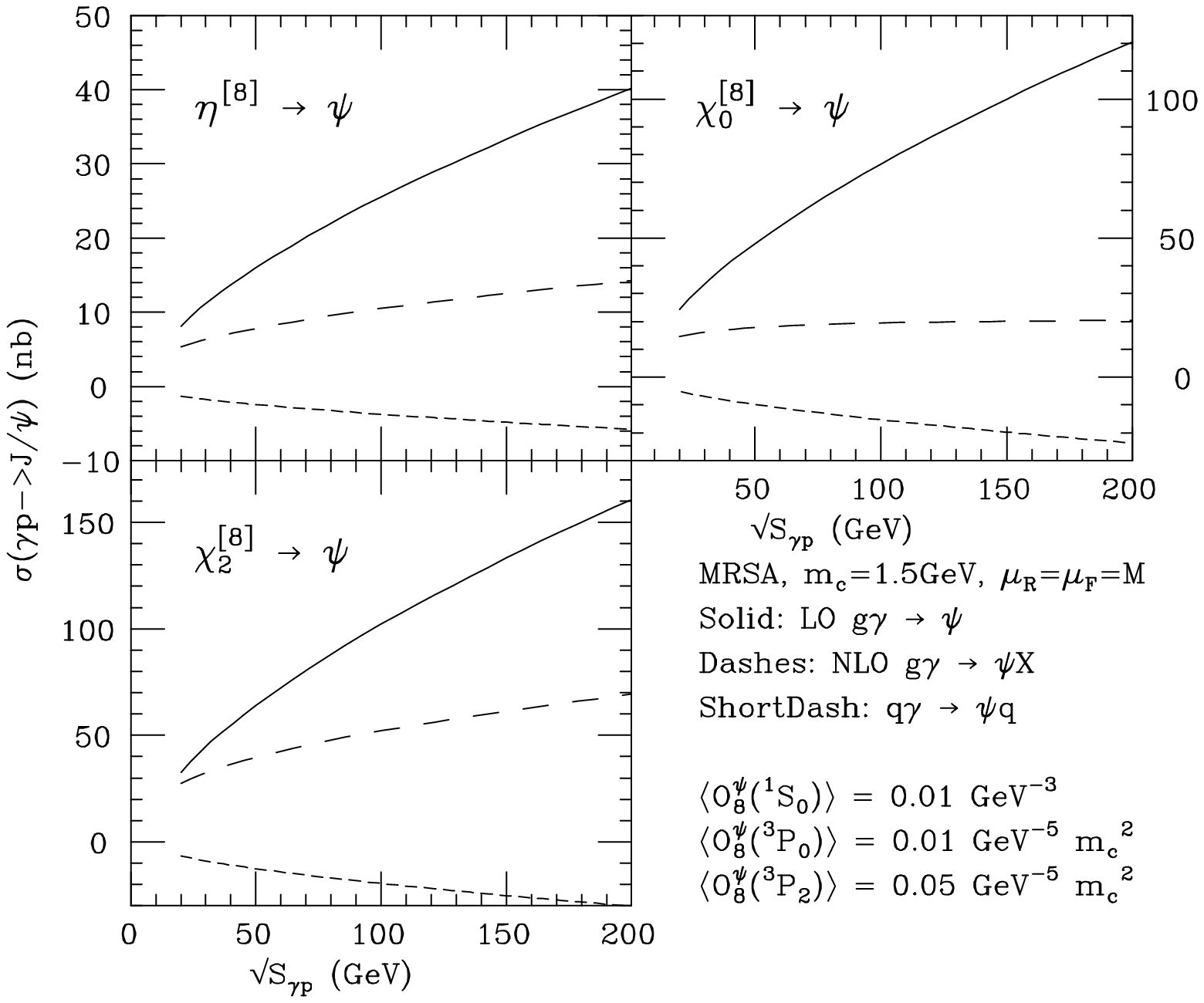,width=0.70\textwidth}
\ccaption{}{ \label{fig:qgcol}     
Separate contributions from Born, $\oatwoa$ $\gamma g$ and $\oatwoa$ $\gamma q$ 
processes to
total photoproduction cross-sections as a function of the $\gamma p$ CM
energy.}                                                  
\end{center}
\end{figure} 
\begin{figure}
\begin{center}
\epsfig{figure=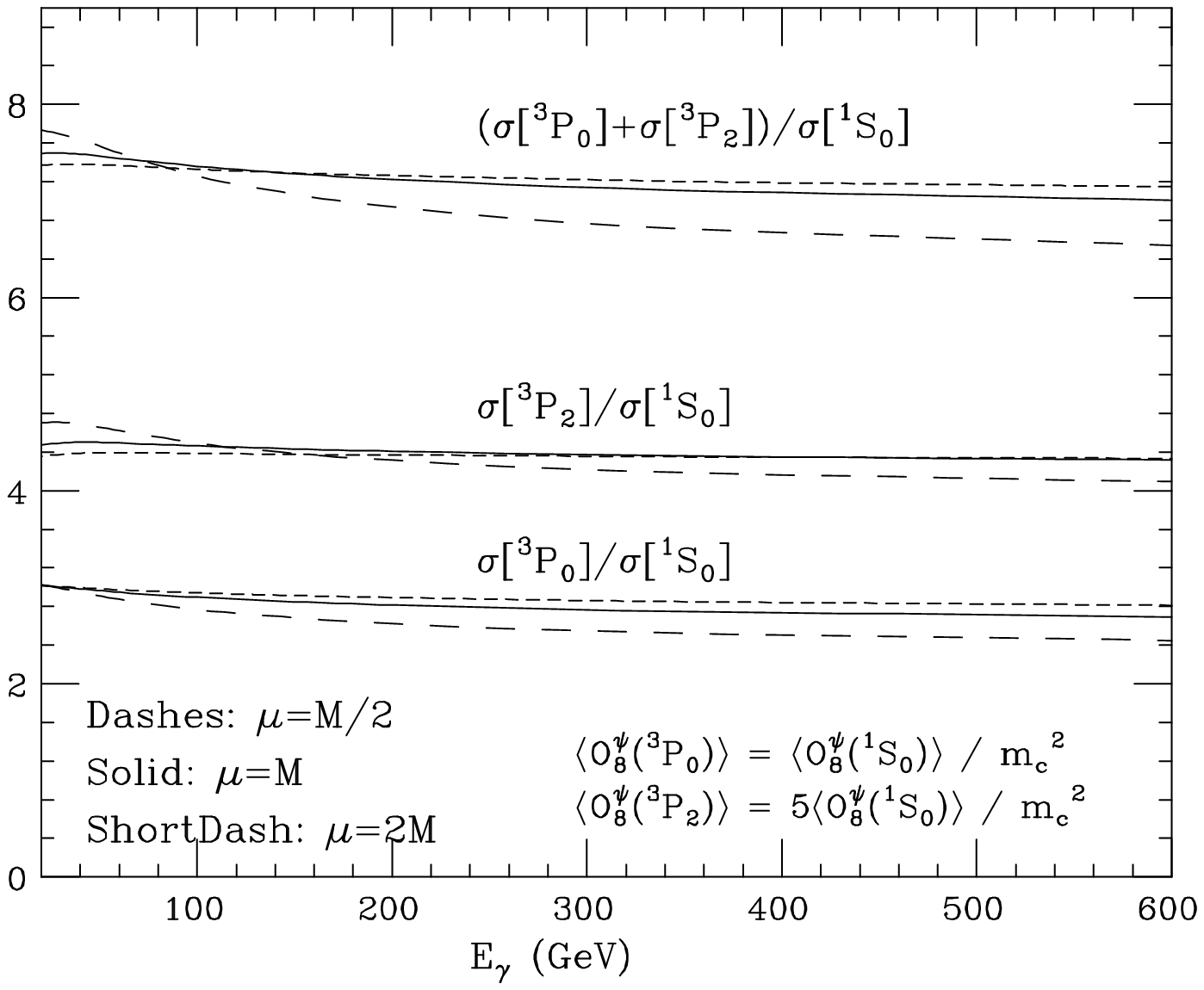,width=0.70\textwidth}
\ccaption{}{ \label{fig:ratfxt}                                    
Ratios of the $\chijh$ to $\etah$ production rates at NLO, normalized to
a common choice of non-perturbative matrix elements,
as a function of the photon beam energy. The Born-level predictions for
these ratios are equal to 3 and 4 for the ratios of $\chizh/\etah$ and 
$\chith/\etah$, respectively.}                            
\end{center}                 
\end{figure} }
The scale dependence
of the results is displayed by the curves relative to the scale choices
$\mur=\muf=aM$, with $a=1/2,1$ and 2. We chose the MRSA set of parton
densities~\cite{mrsa}, including the low-$Q^2$ corrections~\cite{mrsaq2}
necessary to
evaluate consistently the cross-sections when using the smaller choice of
scales, $\muf=M/2$. These low-$Q^2$ corrections are essential to properly
estimate the true scale-dependence of the calculation. Similar curves for the
energy range typical of the HERA collider (given as a function of 
$\sqrt{S_{\gamma p}}$, the $\gamma
p$ CM energy), are shown in fig.~\ref{fig:totcol}. Notice the significant
improvement in the scale dependence when the NLO corrections are included. This
improvement is less remarkable at low energies. The reason for this is that
the $q\gamma \to q \q$ is particularly important at low
energies. This process first appears at $\oatwoa$, and is therefore calculated
at LO only. Its contribution becomes less important at high energies, where
the gluon-initiated process dominates, and the $\oatwoa$ calculations are
therefore genuinely NLO. As a result of this, the $K$-factor (defined as the
ratio of the $\oatwoa/$Born results) is larger at low energy (exceeding a
factor of 2) than at high energies.

The relative importance of the gluon and quark processes is displayed in
figs.~\ref{fig:qgfxt} and \ref{fig:qgcol}, for the fixed-target and 
collider-energy ranges, respectively. Notice that the contribution of the
$q\gamma$ channel is always negative. This is the result of the subtraction of
the mass singularities present in the $q\gamma \to q \q$ channel when the
final-state quark is emitted collinear to the beam. The collinear singularities
are absorbed in the gluon parton density at NLO, as explained in the previous
sections. The net result of this subtraction, in the \MSB\ scheme, is a
negative contribution of the factorized $q\gamma$ process.
                                                          
An important element in previous Born-level 
extractions~\cite{Cacciari96,gammap}
of the non-perturbative colour-octet
parameters from the fit to fixed-target and HERA data is the 
set of relations:
\be
    \sigma^{\psi}[\chizh] \=
     3\frac{m_c^2 \opchizh}{\opetah}\sigma^{\psi}[\etah] 
\quad , \quad
    \sigma^{\psi}[\chith] \=
     4\frac{m_c^2 \opchizh}{\opetah}\sigma^{\psi}[\etah] \; .
\ee                                                          
In these relations we assumed, as usual, $\opchijh=(2J+1)\opchizh$.
As a result of these relations, one could use the following identity:
\be        \label{eq:ratios}
     \sigma(\psi) = \Theta \; \hat{\sigma}[\etah] \;, \quad\quad 
     \mbox{with } \quad                   
     \Theta=\opetah+7\frac{ \opchizh}{m_c^2} \; .       
\ee
The validity of these Born-level relations when the NLO corrections are
included is studied in figs~\ref{fig:ratfxt} and \ref{fig:ratcol}, which show
the ratios of $\chijh$ and $\etah$ production at NLO. As the figures indicate,
the relation in eq.~(\ref{eq:ratios}) holds at NLO to within 10\%. Notice that
the individual relative contributions of the $\chizh$ and $\chith$ states can
however change by up to 30\% with respect to the Born-level prediction 
$\sigma[\chizh]/\sigma[\chith]$ = 4/3.

\begin{figure}
\begin{center}
\epsfig{figure=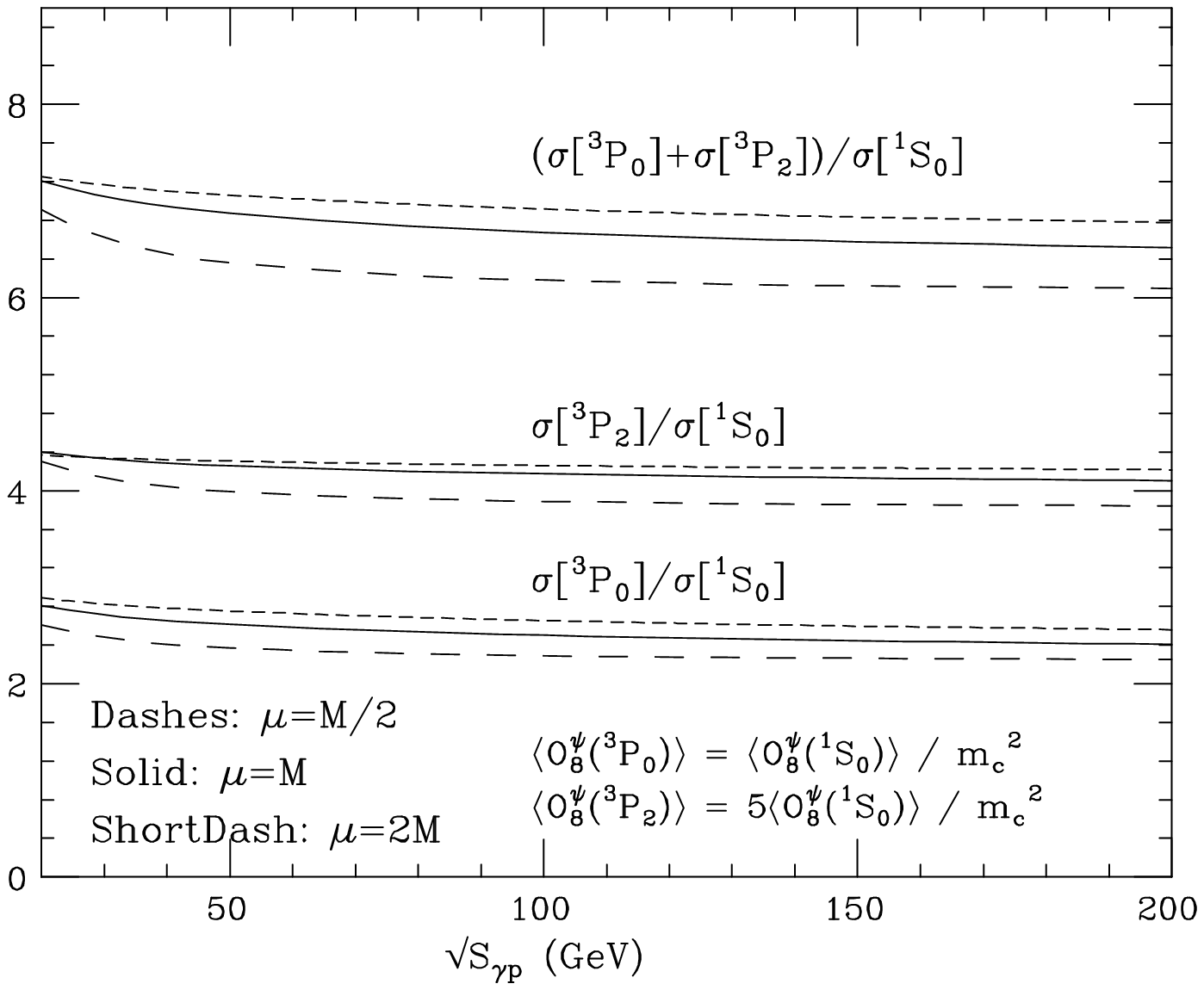,width=0.70\textwidth}
\ccaption{}{ \label{fig:ratcol}     
Same as fig.~\ref{fig:ratfxt}, 
as a function of the $\gamma$p CM
energy. }
\end{center}                 
\end{figure}

\subsection{High-energy behaviour}
While the current experiments only allow to study $\psi$ photoproduction up to
CM energies of the order of few hundred GeV, it is interesting to consider the
behaviour of the total cross-sections in the asymptotic regime. In this regime,
interesting phenomena are expected to take place, because of the 
potentially large small-$x$ effects associated to the presence of diagrams with
$t$-channel gluon exchange. Because of these contributions, the total partonic
cross-sections tend to a constant limit when $x\to 0$. It is easy to see, in
fact, that:
\be  \label{eq:highe}
     \dsh[\gamma \,i \to \q i]\stackrel{x \to 0}{\to}
       - \, \left(\frac{\as}{\pi}\right) \;Z_i \; Z_{\q} \; 
        \szh[\gamma g\to \q] \; , \quad (i=q,g) \; ,     
\ee                                                                
where
\be
    Z_q=\cf \quad, \quad Z_g=\ca \; ,
\ee                          
and
\ba                                   
\label{eq:etahe}                        
       Z_{\etah} &=& 1-\log(\frac{M^2}{\muf^2} )\; , \\
\label{eq:chi0he}
       Z_{\chizh} &=& \frac{43}{27} - \log(\frac{M^2}{\muf^2})    \; ,\\
\label{eq:chi2he}
       Z_{\chizh} &=& \frac{53}{36} -\log(\frac{M^2}{\muf^2})  \; .
\ea                                                                     
Notice that in the $x\to 0$ limit the ratio between the quark and gluon
channels is equal to $\cf/\ca$, as expected.  The factors for the $\gamma q$
channel are the same as those for the  $g q$ channel~\cite{Petrelli97}. The
factors for the $gamma g$ channel are half  those for the $gg$
channel~\cite{Petrelli97}. This is as expected, since  in this last case the
soft $t$-channel gluon can be radiated from either of the two initial state
gluons. These results are trivial but useful cross-checks of our calculations.

Notice also that the small-$x$ partonic cross-sections tend to a negative
value, unless the factorization scale is chosen to be very small.  The large
contribution coming from the small-$x$ region, and the large scale-dependence
of the asymptotic $x\to 0$ limit, suggest that the NLO calculations should
display a large dependence on the scale and on the shape of gluon densities at
sufficiently high CM energies.  A similar behaviour has already been observed
in the NLO hadroproduction case~\cite{Mangano96}.
                                                    
To study this issue in more detail, we assume the gluon density to take, at a
given scale $\muf$, the form:     
\be                     
	G(x)=\frac{1}{x^{1+\delta}} \quad , \quad 0<\delta<1\; .
\ee                                              
The softer the gluon density ($\delta\to0$), the more important the
small-$x$ contributions will be. In the extreme case of $\delta=0$, 
it is easy to find the following result for the NLO/LO $K$-factors of
the total production cross setions, exact up to order $\rho=M^2/S_{\gamma p}$:
\be
K[\q] \= 1+ \frac{\as}{\pi} \left[ \frac{1}{2} A_{\rm tot}[\q] 
       +\ca k_{\q} - \frac{11}{6} \ca\log (\frac{M^2}{\muf^2}) 
   - \ca Z_{\q} \log(\frac{1}{\rho})\right]                   
\ee      
where the coefficients $A_{\rm tot}[\q]$ are collected in
Appendix~\ref{appNLO}, the $Z_{\q}$'s are given in
eqs.~(\ref{eq:etahe})--(\ref{eq:chi2he}) and
\ba                                   
k_{\etah} &=&    
 \frac{25}{3} - \frac{13}{24}\pi^2          
\\                                                                          
k_{\chizh} &=&        
\frac{43}{3} -\frac{467}{432}\pi^2            
\\                                                                
k_{\chith} &=&        
\frac{595}{48} - \frac{1067}{1152}\pi^2 
\ea                                                                        

{
\nopagebreak
\begin{figure}
\begin{center}
\epsfig{figure=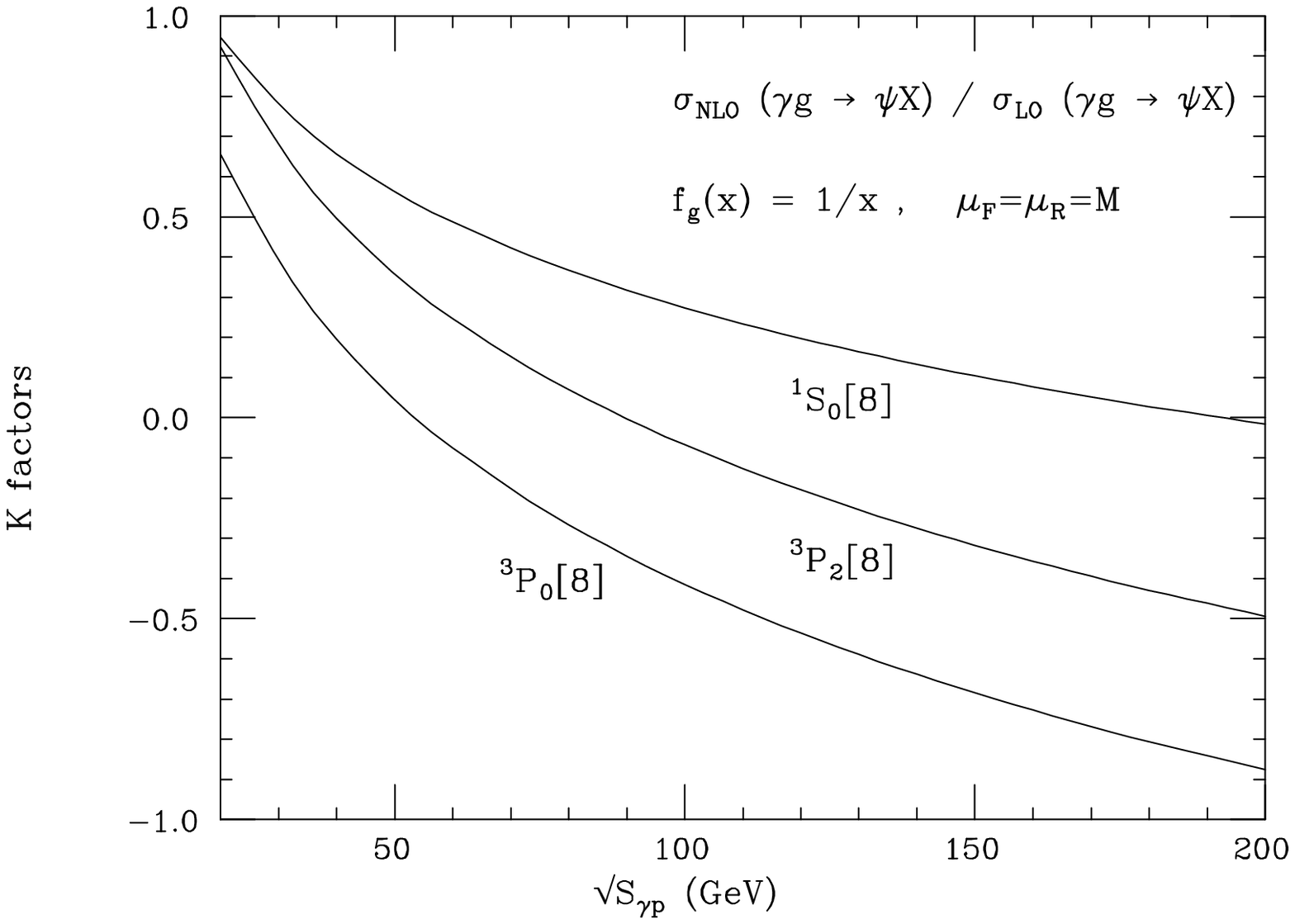,width=0.700\textwidth}
\ccaption{}{ \label{fig:smallx}     
High-energy behaviour of the $K$ factor for the $\gamma g$ process,
using a gluon density $G(x)=1/x$, as a function of the $\gamma$p CM
energy. }                         
\end{center}                 
\end{figure}
\begin{figure}
\begin{center}
\epsfig{figure=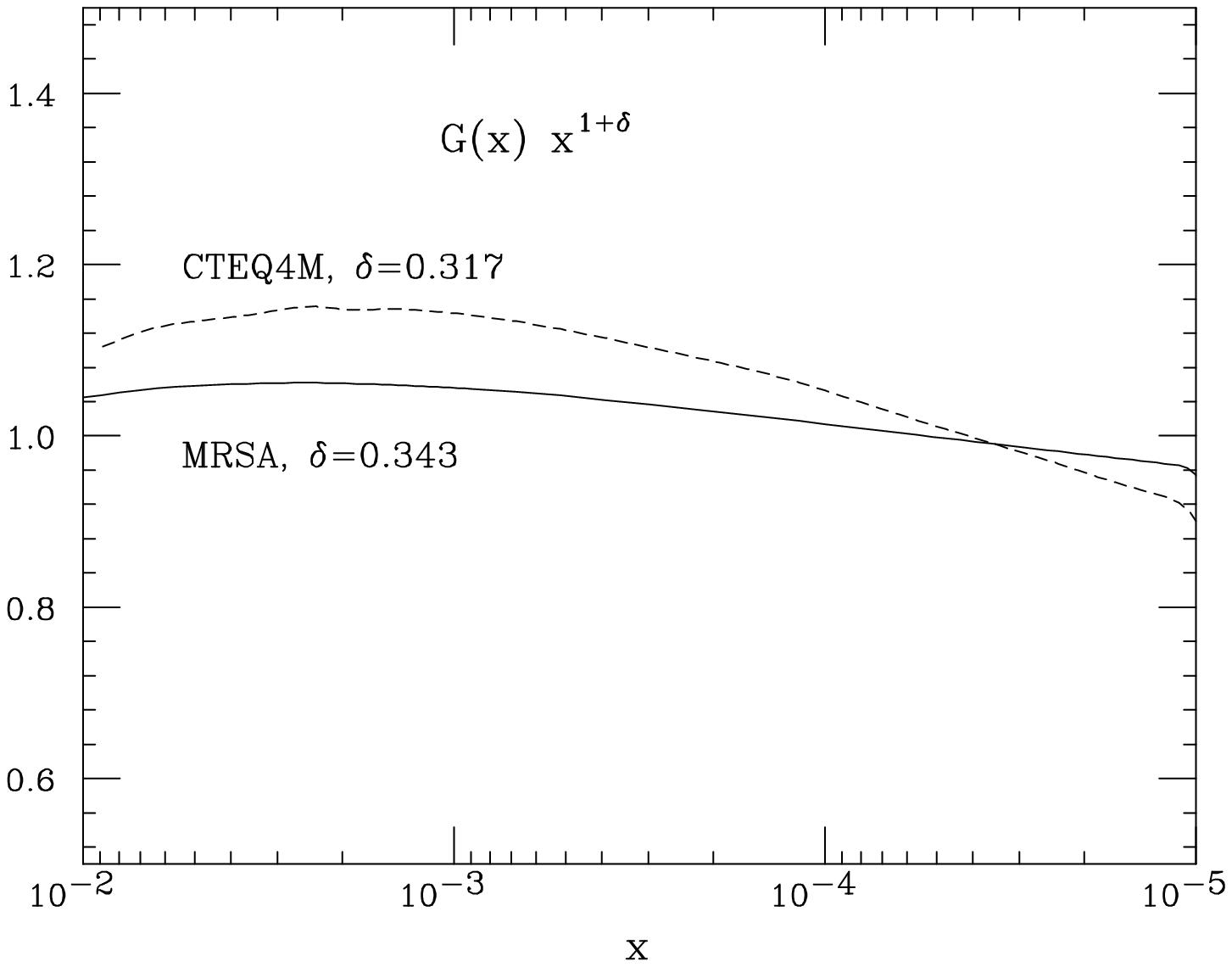,width=0.700\textwidth}
\ccaption{}{ \label{fig:gxfit}     
Small-$x$ behaviour of the gluon density at $\muf=3$~GeV, for the
MRSA~\cite{mrsa} and CTEQ4M~\cite{Lai97} fits. Arbitrary normalization.}
\end{center}                                                            
\end{figure} 
}
As anticipated, large negative logarithmic terms arise.
The scale
dependence of the coefficient of the $\log\rho$ terms cannot be used to change
the overall sign, since it is formally compensated by the scale dependence of
the gluon density, which in the $\delta=0$ case is given by:
\be                                             
    \frac{dG(\rho)}{d\log{\mufsq}}=\frac{\as}{2\pi}\; \int_{\rho}^{1} \;
   \frac{dz}{z} \; P_{gg}(z) \; G(\rho/z,\muf) \sim \frac{\ca\as}{\pi}G(\rho)
   \log\frac{1}{\rho}.
\ee                                                                      
The behaviour of these functions is shown in fig.~\ref{fig:smallx}. The 
$K$-factor becomes negative already for relatively small values of  
$\sqrt{S_{\gamma p}}$, making the NLO estimates unreliable. 

\begin{figure}
\begin{center}
\epsfig{figure=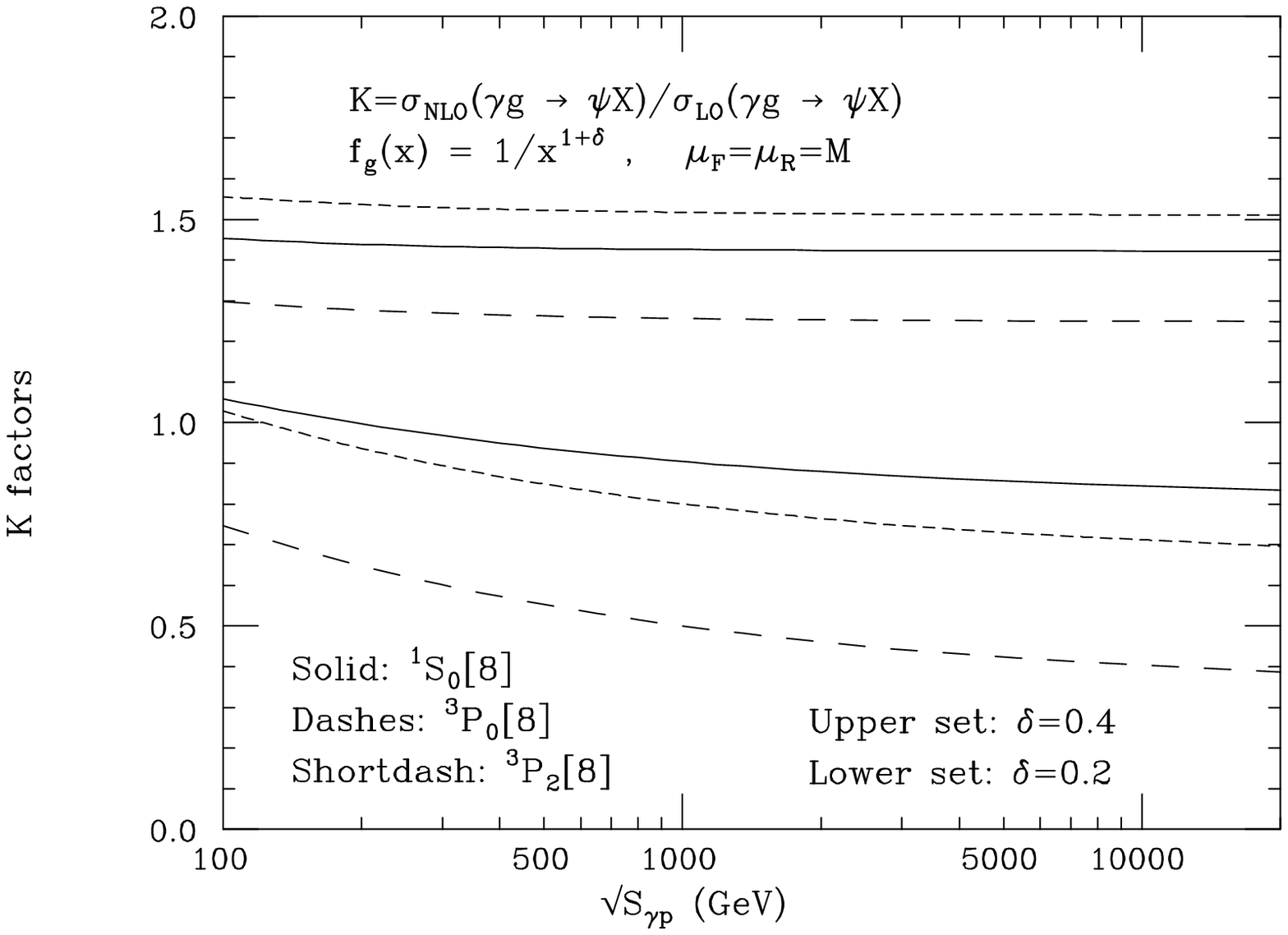,width=0.700\textwidth}
\ccaption{}{ \label{fig:smallx2}     
High-energy behaviour of the $K$ factor for the $\gamma g$ process,
using a gluon density $G(x)=1/x^{1+\delta}$, as a function of the $\gamma$p CM
energy and for different values of $\delta$.}                                 
\end{center}                                
\end{figure} 
This result is not inconsistent with the nice behaviour of the cross-sections
at $\sqrt{S_{\gamma p}}\sim 100$~GeV seen in fig.~\ref{fig:totcol}, since in
that figure the results were obtained using the more realistic set of parton
densities MRSA. 
Typical values of $\delta$ at $\muf=3$~GeV for the current
fits of the gluon densities~\cite{mrsa,Lai97} 
are around $\delta=0.3$ (see fig.~\ref{fig:gxfit}).
The stability of the predictions in fact improves for $\delta >0$, when the
contribution of the small-$x$ regime is suppressed. 
One can easily verify in fact that for $\delta \log(1/\rho)\gg 1$
the asymptotic behaviour of the $K$-factor is given by:
\be   \label{eq:Ksteep}
	K[\q] \stackrel{\rho \to 0}{\to} 1+ \frac{\as}{\pi} 
        \frac{1}{\delta} \left(
         Z_{\q} +
          {\cal O}(\delta,\rho) \right) \; ,
\ee                                      
Depending on the value of $\delta$, the NLO correction  can still be large and
negative. 
Figure~\ref{fig:smallx2} shows the high-energy behaviour of the $K$-factor for
the values $\delta=0.2$ and $\delta=0.4$ \footnote{We chose not to use directly
current fits of the gluon densities, since in any case these are not reliable
in the range $x<10^{-5}$ ($\sqrt{S_{\gamma p}} \gsim 1$~TeV).}.
As predicted by eq.~(\ref{eq:Ksteep}), the $K$-factor tends to a constant.
                                                             
\section{Conclusions}
We presented in this paper the first calculation of $\oatwoa$ corrections to
the total quarkonium photoproduction cross-sections. The calculations have been
performed in the framework of the NRQCD approach to quarkonium production. As
in the Born-level case, the production is dominated by 
colour-octet $\QQ$ states. The contribution to         
the cross-sections from the NLO corrections is large at typical fixed-target 
energies (up to a
factor of 2 increase over the Born results), and decreases at energies
typical of the HERA collider.                                         
In this energy range the NLO corrections significantly improve the stability of
the calculated rates under variations of the renormalization and factorization
scales. 
For energies above few hundred GeV in the $\gamma p$ CM frame, large and
negative small-$x$ contributions dominate the production rate, and make the NLO
evaluation of the total cross-sections strongly dependend on the
scale and shape of the gluon density, calling for the resummation
of small-$x$ effects.               

The calculations presented in this paper do not directly 
improve our knowledge of the quarkonium photoproduction at $\pt>0$, since
production at $\pt>0$ is a process of $\oatwoa$ at the Born level. Nevertheless
they provide a fundamental element in the determination of the range of $\pt$
and $z=p_{\psi}\cdot P_N/p_{\gamma}\cdot P_N$ where the Born calculations can
be reliably used. In fact the large negative contributions arising in the
$\pt=0$ and $z=1$ regions from the virtual corrections evaluated in this paper 
affect, via perturbative Sudakov effects, the estimate of the production rates
near the end-points of the elastic region. 
The implications of these effects in the comparison of the $z$ and $\pt$
distributions measured at HERA with the predictions of QCD will be examined in
a forthcoming publication.

\vspace{1cm}
{\bf Acknowledgments.} We thank M. Cacciari for providing us with the
expression of the real-emission matrix elements~\cite{Cacciari96}
which were used in this work.  One of us (A.P.) thanks the CERN Theory Division
for hospitality while part of this work was being done.
%
%
%
%

\appendix
\section{Symbols and notations}
\label{appA}
This Appendix collects the meaning of various symbols which are used throughout
the paper.
\\[0.2cm]
\underline{Kinematical factors}:
\be                            
M = 2m \; , \quad
v={\sqrt{1-\frac{M^2}{s}}}\; , \quad
\rho= \frac{M^2}{S_{\gamma p}}\; , \quad
\beta=\left(1-\rho\right)^{\frac{1}{2}}\; ,
\ee
where $s$ is the partonic center of mass energy squared and $S_{\gamma p}$ 
is the $\gamma$-hadron
one. $v$ is the velocity of the bound (anti)quark in 
the quarkonium rest frame, $2v$ being then the relative velocity of the quark
and the antiquark. We also define:
\be                                          
\feps{Q^2} = \left(\frac{4 \pi\mu^2}{Q^2}\right)^{\epsilon}\Gamma(1+\ep) 
= 1+\ep\left(-\gamma_E +\log(4\pi) + \log{\mu^2 \over Q^2}\right) +
  {\cal O}(\epsilon^2) \,,
\ee   
and we denote a perturbative $\QQ$ state with generic spin and angular momentum
quantum numbers and in a colour-singlet or colour-octet state
by the symbol                                         
\be
{\cal Q}^{[1,8]}  \equiv \QQ[\spectr^{[1,8]}] \; .
\ee                                               
\\[0.2cm]
\underline{Altarelli-Parisi splitting functions}. Several functions related to
the AP splitting kernels enter in our calculations. We collect here our
definitions:
\ba                                             
&&P_{q\gamma}(x) = \df \left[x^2+(1-x)^2-2\ep\ x(1-x)\right]\\
&&{\cal P}_{q\gamma}(x) = \df \left[x^2+(1-x)^2\right]\\
&&P_{gq}(x)=\cf\left[\frac{1+(1-x)^2}{x}-\ep\ x\right]\\
&&{\cal P}_{gq}(x)=\cf\left[\frac{1+(1-x)^2}{x}\right]\\
&&P_{gg}(x) = 2\ca\left[\frac{x}{1-x}+\frac{1-x}{x}+x(1-x)\right]\\
&&{\overline P}_{gg}(x) = 
2\ca\left[\frac{x}{(1-x)_{\rho}}+\frac{1-x}{x}+x(1-x)\right] \\
&&{\cal P}_{gg}(x) = {\overline P}_{gg}(x) + \left(\b0+4
\ca\log\beta\right)\delta(1-x)
\ea
where 
\be           
\b0 = \frac{11}{6}\ca- \frac{2}{3} \tf\nf
\ee
with $\nf$ the number of flavours {\sl{lighter}} than the bound one.  
The $P_{ij}$ are the $D$-dimensional splitting
functions which appear in the factorization of collinear singularities from
real emission,while the functions ${\cal P}_{ij}$ are the four-dimensional AP
kernels, which enter in the \MSB\ collinear counter-terms.
The $\rho$-distributions are defined by:                  
\be                                                                       
   \int_{\rho}^{1} \, dx \, \left[d(x)\right]_{\rho} t(x) \=
   \int_{\rho}^{1} \, dx \; d(x) \; \left[t(x)-t(1)\right] \; .
\ee                                                   
\\[0.2cm]
\underline{Colour Algebra}
\ba
[T^a, T^b] &=& i f^{abc} T^c \\
\{T^a, T^b\} &=& d^{abc} T^c + \frac{\delta^{ab}}{\nc}
\ea
\be
\begin{array}{ll}
 \tr(T^a T^b) = \tf \delta^{ab} &\tf = \frac{1}{2}\\
\sum_a (T^a T^a)_{ij} = \cf \delta_{ij}
&\cf = \frac{N_c^2-1}{2N_c} = \frac{4}{3}\\
\sum_{bc} f^{abc} f^{ebc} = \ca \delta^{ae} &\ca = \nc = 3 \\
\sum_{abc} d^{abc} (T^a T^b T^c)_{ij} = C_2(F) \delta_{ij}
&C_2(F) = \frac{(N_c^2-4) (N_c^2-1)}{4 N_c^2} = \frac{10}{9} \\
\end{array}
\ee
\ba
\df &=& \sum_i \delta_{ii} = \nc = 3 \\
\da &=& \sum_a \delta^{aa} = N_c^2 -1 =8 
\ea
The following formulas were  found to be useful:
\ba           
\sum_a T^a_{ij} T^a_{kl}   &=& \frac{1}{2}
\left( \delta_{il} \delta_{jk} - 
\frac{1}{\nc} \delta_{ij} \delta_{kl} \right)\\ 
\tr (T^a T^b T^c) &=& \frac{1}{4} ( d^{abc} + i f^{abc} ) \\
\tr (T^a \{T^b ,T^c\})&=& \frac{1}{2} d^{abc}\\
\cf \df &=& \tf \da 
\ea                 
\\[0.2cm]
Notice that, according to the discussion in
ref.~\cite{Petrelli97},
our conventions differ slightly from those introduced in
ref.~\cite{bbl} (and labelled here as BBL):
\ba                                       
\label{eq:opnorm}
&&\langle {\o}_1\rangle = \frac{\langle {\o }_1\rangle ^{\rm BBL}}{2\nc}, \\
&&\langle {\o}_8\rangle = \langle {\o }_8\rangle ^{\rm BBL}. 
\ea

\subsection{$O(\as \aem)$ cross-sections}
The $D$-dimensional cross-sections read
\ba
&&\sigma(ij\to\spectr^{[1,8]}\to H) = \hat \sigma(ij\to\spectr^{[1,8]}) 
{{\langle{\o }_{[1,8]}^H(\spectr)\rangle}
\over{N_{col} N_{pol}}},
\ea
the short distance coefficients $\hat\sigma$  having been 
calculated according to the rules of Section~\ref{sec:projectors}.
$N_{col}$ and $N_{pol}$ refer to the number of colours and polarization states
of the $\QQ[^{2S+1}L_J^{[1,8]}]$ pair produced. They are given by 1 for 
singlet states or $\da = N_c^2-1$ for octet states, 
and by the $D$-dimensional $N_J$'s defined
in Section~\ref{sec:projectors}.
Recall that  the matrix elements appearing in the equations
above are meant to be the bare $D$-dimensional ones. 
Making use of their correct
mass-dimension, $3-2\ep$ and $5-2\ep$ for $S$ and $P$ wave states 
respectively, gives the right dimensionality
to $D$-dimensional cross-sections, i.e. $2-D = -2+2\ep$.

We shall use the short-hand notation
\be
\sigma^H(i j  \to \q^{[1,8]} ) \equiv \sigma(i j  \to {\q}^{[1,8]} \to H) 
\ee                                        
to indicate the production process  of the physical quarkonium state
$H$ via the intermediate $\QQ$ state ${\q}^{[1,8]} = \QQ[\spectr^{[1,8]}]$.

The $D$-dimensional Born cross-sections read:
\ba 
&&\sbh(g\gamma \to \oneSzero^{[8]}) = 
{{2\as \aem \ech\mu^{4\ep}\pi^3}\over{ m^5}}
{{1-2\ep}\over{1-\ep}} \delta(1-x) 
{{\langle {\o}^H_8(\oneSzero)\rangle}\over{\da}} \\&\nn\\
&&\sbh(\qq \to \threeSone^{[8]}) = \frac{\da}{{\df}^2}{{\assq\mu^{4\ep}\pi^3}\over{2 m^5}}(1-\ep)   \delta(1-x) 
{{\langle {\o}^H_8(\threeSone)\rangle} \over {\da (3-2\ep)}}\\
&&\sbh(g\gamma \to \threePzero^{[8]}) = {{18\as \aem \ech\mu^{4\ep}\pi^3}\over{ m^7}}
{1\over{(1-\ep)(3-2\ep)}} \delta(1-x) 
{{\langle {\o}^H_8(\threePzero)\rangle}\over{\da}}\\
&&\sbh(g\gamma \to \threePone^{[8]})=0\\
&&\sbh(g\gamma \to \threePtwo^{[8]}) = 
{{4 \as \aem \ech \mu^{4\ep}\pi^3}\over{m^7}}
{{6-13\ep+4\ep^2}\over{(1-\ep)(3-2\ep)}} \delta(1-x) 
{{\langle {\o}^H_8(\threePtwo)\rangle}\over{\da (1-\ep)(5-2\ep)}} 
\ea

\section{Summary of  $O(\assq \aem) $ Results}
\label{appNLO}                 

We define:
\be
\sigma_0^H(\gamma j\to {\cal Q}) \delta(1-x)\equiv   
\sigma_{\rm Born}^H(\gamma j\to {\cal Q}) 
\ee
The $O(\assq \aem)$ cross-sections are 
given as a function of the variable $x=M^2/s$.

\noindent                   
{\bf The $\gamma g \to \q^{[8]} X$  channels}
                                    
\ba
\dsh[\gamma \,g\to\psih\,g] &=& \frac{\assq \aem \ech\pi^2}{{(2 m)}^5}
f_{\gamma g}[\psih](x)\langle\o_8^H(\threeSone)\rangle\\
f_{\gamma g}[\psih](x)&=& \label{fggpsih}
\frac{80 x^2}{9 (-1 + x)^2 (1 + x)^3}
\left[ 2 + x + 2 x^2 - 4 x^4 - x^5 \right. \nn\\
&&\left. + ( 10 x^2  + 4 x^3  + 2 x^4) \log x \right]\\
&&\nn\\
\dsh[\gamma \,g\to\chioh\,g] &=& \frac{\assq \aem \ech \pi^2}{{(2 m)}^7}
f_{\gamma g}[\chioh](x)\langle\o_8^H(\threePone)\rangle\\
f_{\gamma g}[\chioh](x) &=&
\frac{32}{3 (-1 + x)^4 (1 + x)^5} 
\left[ 5 - 4 x - 71 x^2 + 26 x^3 - 67 x^4 + 75 x^5 \right. \nn\\
&&\left. + 247 x^6 - 23 x^7 - 82 x^8 - 
    71 x^9 - 32 x^{10} - 3 x^{11} + ( - 18 x^2  - 6 x^3  \right.\nn\\
&&\left. - 270 x^4  + 138 x^5  - 6 x^6  + 186 x^7  + 
    186 x^8  + 66 x^9  + 12 x^{10}) \log x\right]
\\
&&\nn\\
\dsh[\gamma \,g\to \q^{[8]}\,g] &=& \szh[\gamma g\to\q^{[8]}]\left(\delta(1-x) 
+ \frac{\as}{2 \pi}
\left\{  \Atot [\q^{[8]} ] \, \delta(1-x) \right.\right.\nn\\
&&+ \left[ x {\overline P}_{gg}(x)\log\frac{4m^2}{x\qf} + 2 x(1-x)
P_{gg}(x)\left(\frac{\log(1-x)}{1-x}\right)_{\rho} +\right. \nn\\
&&+ \left. \left. \left. \left(\frac{1}{1-x}\right)_{\rho} f_{\gamma g}[\q^{[8]}](x)
   \right]\right\} \right)\, , \nn\\[10pt]
&&[\q^{[8]}=\etah,\chizh,\chith],
\ea
where:
\ba
\Atot [\etah]   &=&\cf\,\left(-10 +\frac{\pi^2}{2}  \right)  + 
  \ca\,\left(  5 - \frac{\pi^2}{6}  \right)\nn\\ 
&&- 4\ca\log \beta  + 
     8\ca\log^2 \beta  
 + 2\,{b_0}\,\log {{\mu }\over {\muf}} + 
     8\ca\log\beta\log\frac{2m}{\muf}\\
\Atot [\chizh]  &=&\cf\,\left( -\frac{14}{3}+\frac{\pi^2}{2} \right)  + 
  \ca\,\left(3 - \frac{\pi^2}{6}    \right)\nn\\ 
&&- 4\ca\log \beta  + 
     8\ca\log^2 \beta  
 + 2\,{b_0}\,\log {{\mu }\over {\muf}} + 
     8\ca\log\beta\log\frac{2m}{\muf}\\
\Atot [\chith] &=&-8\cf\,+ 
  \ca\,\left(  \frac{7}{2}+ \frac{\pi^2}{3}+ \log 2  \right)\nn\\ 
&&- 4\ca\log \beta  + 
     8\ca\log^2 \beta  
 + 2\,{b_0}\,\log {{\mu }\over {\muf}} + 
     8\ca\log\beta\log\frac{2m}{\muf}\\
f_{\gamma g}[\etah](x)&=&
\frac{2\ca}{(1 - x) (1 + x)^3} 
\left[-1 - x^2 - 2 x^3 + 2 x^5 + x^6 + x^8 \right.\nn\\
 &&\left.+(1 - 4 x^2  - 2 x^4  - 4 x^6  + x^8 ) \log x \right]\\
f_{\gamma g}[\chizh](x)&=&
\frac{2\ca}{27 (1 - x)^3 (1 + x)^5}
\left[-43 + 14 x + 193 x^2 - 82 x^3 - 457 x^4 + 342 x^5 \right.\nn\\
&& \left. + 124 x^6 - 302 x^7 + 
      113 x^8 + 28 x^9 + 43 x^{10} + 27 x^{12} + (27   - 90 x^2   + 
   42 x^3 \right. \nn\\
&&\left.   + 135 x^4   + 78 x^5   - 750 x^6   + 
   270 x^7   - 117 x^8   - 6 x^9   - 192 x^{10}   + 
   27 x^{12}) \log x \right]\\
f_{\gamma g}[\chith](x)&=& 
\frac{\ca}{18 (1 - x)^3 (1 + x)^5}
\left[-53 + 16 x + 119 x^2 - 86 x^3 - 521 x^4 + 621 x^5 - 55 x^6\right.\nn\\ 
&&\left. 
- 97 x^7 + 394 x^8 - 445 x^9 + 80 x^{10} - 9 x^{11} + 36 x^{12} + 
( 36 - 126 x^2   + 66 x^3  - 378 x^4   \right.\nn\\
&&\left.   + 210 x^5  - 798 x^6   + 1242 x^7  
   - 810 x^8   + 402 x^9   - 168 x^{10}   + 36 x^{12}) \log x \right]
\ea
\noindent

\newpage
{\bf The $\gamma q(\bar q) \to \q^{[8]} X$  channels}
                                            
\ba
\dsh[\gamma\,q\to \psih\,q]&=& \frac{\aem \ech}{\pi}\szh[\qq\to\psih]  \nn\\
&&\times\left\{\left[\frac{x}{2}
P_{q \gamma }(x)\log\frac{4m^2(1-x)^2}{x\qf} + \df
x^2(1-x)\right]+ f_{\gamma q}[\psih](x)\right\}\\ 
&&\nn\\
&&\nn\\
\dsh[\gamma\,q\to \chioh\,q] &=& \frac{\pi^2\assq \aem \ech}{{(2 m)}^7}
f_{\gamma q}[\chioh](x)\langle\o_8^H(\threePone)\rangle\\
&&\nn\\
&&\nn\\
\dsh[\gamma\,q\to \q^{[8]}\,q] &=& \frac{\as}{\pi}\szh[\gamma g\to\q^{[8]}] 
\times \nn\\
&&\left\{ 
\left[
\frac{x}{2} P_{gq}(x)\log\frac{4m^2(1-x)^2}{x\qf} + \cf\frac{x^2}{2}
\right]             
+ f_{\gamma q}[\q^{[8]} ](x)\right\},\\[10pt]
&&[\q^{[8]}=\etah,\chizh,\chith]\nn    
\ea
where
\ba
&&f_{\gamma q}[\psih](x)= - \frac{\df }{4} x(- 1 + x)  (1 + 3 x)
\label{fgqpsih}\\
&&f_{\gamma q}[\chioh](x)=\frac{1}{\df} \frac{128}{3}
\left[\frac{1}{3} (-1 + x) (-5 + 4 x + 4 x^2 ) -  x^2 \log x \right]
\\
&&f_{\gamma q}[\etah](x)= \cf(-1+x)(1-\log x) \\
&&f_{\gamma q}[\chizh](x)=\frac{\cf}{9}\left[\frac{1}{3}(-1+x)(43-14 x+4
  x^2) + (9-9x+4x^2)\log x\right]\\
&&f_{\gamma q}[\chith](x)=\frac{\cf}{12}\left[\frac{1}{3}(-1+x)(53-16 x+20
  x^2)+(12-12x+5x^2)\log x\right]
\ea                              

\end{document}